\documentclass[aps,prb,reprint,amsmath,amssymb]{revtex4-1}

\usepackage{graphicx}
\usepackage{modiagram}
\usepackage{braket}
\usepackage{ulem}   
\usepackage{dsfont}
\usepackage{amsthm,amsmath,amsfonts,amssymb,verbatim,color,tabularx,adjustbox}
\usepackage{bbold}
\usepackage{mathtools}
\usepackage[T1]{fontenc}
\usepackage[colorlinks=true,citecolor=blue,linkcolor=blue,urlcolor=blue]{hyperref}
\newcommand{\defect}[2]{$\mathrm{{#1}}_{\mathrm{{#2}}}$}

\begin{document}

\title{Electronic and Structural Properties of Lanthanide-Doped MoS$_2$: Impact of Ionic Size and Orbital Configuration Mismatch}

\author{Hyosik Kang}
\affiliation{Department of Chemistry, Pennsylvania State University, University Park, Pennsylvania 16802, USA}
\author{Raquel Queiroz}
\affiliation{Department of Physics, Columbia University, New York, NY 10027, USA}
\author{Lukas Muechler}
\affiliation{Department of Chemistry, Pennsylvania State University, University Park, Pennsylvania 16802, USA}
\affiliation{Department of Physics, Pennsylvania State University, University Park, Pennsylvania 16802, USA}
\email{lfm5572@psu.edu}

\begin{abstract}
Single-photon emitters (SPEs) are crucial for quantum technologies such as quantum simulation, secure quantum communication, and precision measurements. Two-dimensional transition metal dichalcogenides (TMDCs) are promising SPE candidates due to their atomically thin nature and efficient photon extraction. However, their emission wavelengths limit compatibility with existing telecommunication technologies. Lanthanide doping in TMDCs, such as \defect{MoS}{2}, offers a potential solution by introducing sharp, $f$-orbital derived emissions in the infrared range. Yet, the feasibility of introducing these dopants remains uncertain due to their large ionic radii of the lanthanides.
We employ density functional theory (DFT) calculations to investigate the structural and electronic properties of lanthanide-doped \defect{MoS}{2} monolayers (Ln=Ce, Er). By evaluating formation energies with up to three adjacent S vacancies, we assess how these vacancies mitigate lattice strain caused by the size mismatch of Ce and Er with Mo. Our results show that while \defect{Ln}{Mo} destabilizes the pristine lattice, S vacancies enhance thermodynamic stability. Charge state analysis indicates that defect states introduced by \defect{Ln}{Mo} localize near the valence band and remain stable across a wide Fermi energy range.
Electronic structure analysis shows that Ce$^{4+}$ and Er$^{3+}$ maintain their oxidation states upon electron doping due to additional acceptor states from host-induced dangling bonds. These states arise from an orbital filling mismatch between dopants and Mo. Consequently, \defect{Ce}{Mo} is unlikely to exhibit infrared emissions due to its empty $f$-shell, whereas \defect{Er}{Mo} is expected to emit in the infrared. These findings demonstrate the potential of lanthanide-doped TMDCs as tunable SPEs and provide design strategies for optimizing their optical and electronic properties.
\end{abstract}
\date\today
\maketitle

\section{Introduction}
Distinct from coherent and thermal light sources, single-photon emitters (SPEs) emit light as uncorrelated single-photons. They play a central role in numerous quantum technologies, such as quantum simulation \cite{SPEforquantum1, SPEforquantum2, SPEforquantum3}, precision measurements \cite{SPEforquantum4}, and secure quantum key distributions \cite{SPEforquantum5, SPEforquantum6}. To achieve single-photon emission, SPEs require high quantum yields and single-photon purity.

Among the most promising materials for single-photon generation are solid-state systems based on point defects, which combine the optical properties of localized defects with the practical advantages of a solid-state host. \cite{solidSPE1,solidSPE2,solidSPE3,solidSPE4}  The optical properties of these localized defects are characterized sharp emission lines with a strong zero-phonon line (ZPL), enabling high quantum yields and single-photon purity which are essential for SPEs. \cite{pointdefectSPE1,pointdefectSPE2,pointdefectSPE3,pointdefectSPE4,pointdefectSPE5,pointdefectSPE6}

The negatively charged nitrogen-vacancy (NV-) center in diamond has been widely regarded as a prime candidate material for SPEs. \cite{pointdefectSPE3, NV-SPE1, NV-SPE2, NV-SPE3, NV-SPE4, NV-SPE5} This is because it possesses bright luminescence, long spin coherence times, and exceptionally sharp emission line corresponding to a ZPL at $1.945 eV$. However, as a three-dimensional material, diamond presents challenges for modifying point defects and efficiently extracting photons from its high-refractive-index matrix, limiting its applicability in many technological applications.

Two-dimensional materials, specifically transition metal dichalcogenides (TMDCs), have been proposed as an alternative platform for SPEs. \cite{TMDCSPE1, TMDCSPE3, TMDCSPE4, TMDCSPE5, TMDCSPE6, TMDCSPE7, TMDCSPE8} The atomically thin nature of TMDCs facilitates efficient photon extraction and integration into devices. However, the limited photon emission range (400-800 nm) of TMDCs constrains their potential applications as SPEs, e.g. in quantum telecommunication devices. 

\begin{figure}[b]
    \centering
    \includegraphics[width=0.5\textwidth]{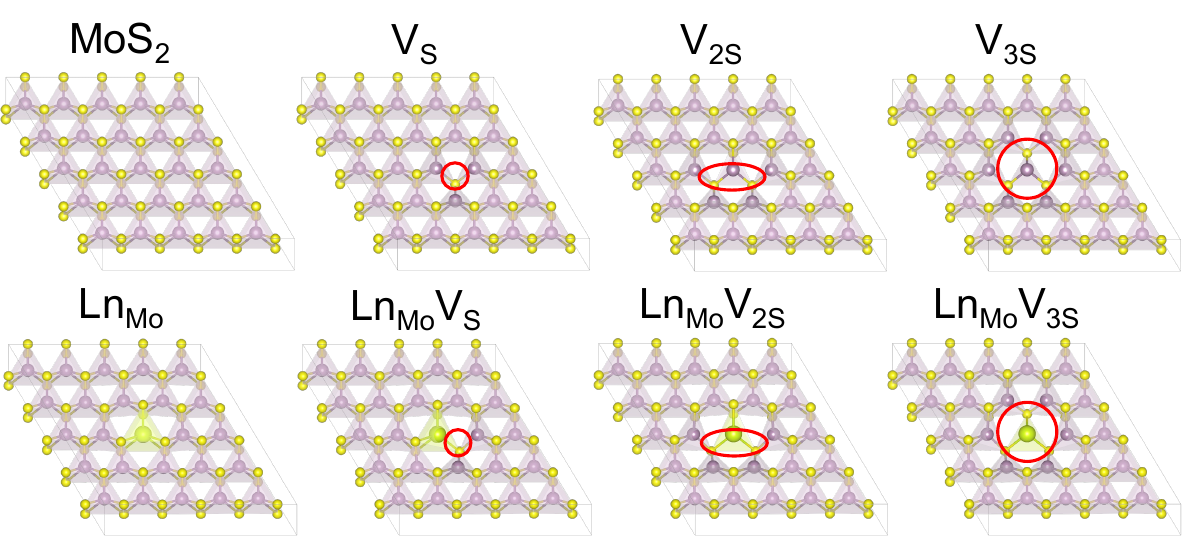}
    \caption{Defect configurations of monolayer \defect{MoS}{2} considered in this paper: single sulfur vacancy (\defect{V}{S}), double sulfur vacancies (\defect{V}{2S}), triple sulfur vacancies (\defect{V}{3S}), lanthanide substituent on the Mo site (\defect{Ln}{Mo}), and \defect{Ln}{Mo} with adjacent sulfur vacancies (\defect{Ln}{Mo}\defect{V}{S}, \defect{Ln}{Mo}\defect{V}{2S}, and \defect{Ln}{Mo}\defect{V}{3S}). The purple, yellow, and green atoms indicate Mo, S, and Ln respectively. Red rings indicate sulfur vacancies.}
    \label{fig_scheme of defect compositions}
\end{figure}

To overcome these limitations, lanthanide dopants have been explored as potential SPEs for these applications, including lanthanide-doped TMDCs. \cite{LnTMDCSPE1, LnTMDCSPE2, LnTMDCSPE3_WS2, LnTMDCSPE4} Unlike $s$-, $p$-, and $d$-orbitals, $f$-orbitals generally do not participate in molecular bonding, resulting in a localized set of orbitals. Furthermore, electron transitions between $f$-orbitals exhibit very sharp emission lines and relatively long lifetimes due to the Laporte rule. Additionally, as seen in erbium-doped fiber amplifiers (EDFAs), lanthanide-doped materials produce emission lines typically in the infrared range with minimal wavelength dispersion, making them highly desirable for telecommunication applications. \cite{Lnusage1, Lnusage2, Lnusage3, Lnusage4} However, lanthanide elements are significantly larger than Mo$^{4+}$ (by at least 20pm \cite{Ionicradii}), suggesting that direct substitution may not be feasible without additional defects or significant lattice distortions. Consequently, atomic-level studies remain challenging in this domain.

In this paper, we investigate the electronic and structural properties of lanthanide-doped (Ln = Ce, Er) \defect{MoS}{2} monolayers using density functional theory (DFT). We calculate the formation energies of these point defects and examine their dependence on adjacent vacancies near the substituent atoms. In addition, we analyze their electronic structures to develop a microscopic understanding of the defects and their properties. \defect{MoS}{2} monolayers are chosen as representative TMDCs. Ce was chosen as a model lanthanide to study $f$-electrons in TMDCs because Ce$^{3+}$ with its nominal $f^1$ represents the simplest case of $f$-electron systems, as most $f$-orbitals are expected to reside in the band gap. Er has been chosen due to its $f-f$ transition band that lies within the telecommunication band range. Up to three adjacent sulfur vacancies (\defect{V}{S}, \defect{V}{2S}, and \defect{V}{3S}) are considered to assess the effect of varying ionic sizes on formation energies (Fig.~\ref{fig_scheme of defect compositions}). This focus is particularly important as sulfur vacancies are among the most common defects in \defect{MoS}{2} monolayers. \cite{VSMoS21_Eform_VMoPCHGD, VSMoS22, VSMoS23_Eform, VSMoS24, VSMoS25_Eform, VSMoS26_Eform, VSMoS27}

\section{Calculation Details}
DFT calculations were performed using the VASP code. \cite{Method_VASP} The generalized gradient approximation (GGA) in the form of the Perdew-Burke-Ernzerhof (PBE) functional and projector-augmented wave (PAW) pseudopotentials were employed \cite{Method_GGA, Method_PAW, Method_PBE, Method_planewavebasisset} The energy cutoff for the plane-wave basis set was set to $500eV$, and the total energy convergence criterion was $10^{-6}$. The self-consistent field (SCF) cycle was spin-polarized to account for the magnetism arising from $f$-electrons. All calculations were performed with fully relaxed atomic positions, and only the $\Gamma$-point was considered. We found that the inclusion of spin-orbit coupling (SOC), does not not have a significant effect on the properties considered here, which is in agreement with previous calculations~\cite{LnTMDCSPE3_WS2}. We therefore only present calculations without SOC.

The formation energy of a defect with charge $q$ is given by 
\begin{equation} \label{eq1}
\begin{split}
    &E_{form}(q) =  \\
    &E_D(q) - E_P + \sum_i \mu_i n_i + q ( E_{VBM} + E_{F} ) + E_{corr},
\end{split}
\end{equation}
where $E_D(q)$ is the energy of the supercell containing the defect with charge $q$, and $E_P$ is the energy of the pristine supercell. The integer $n_i$ denotes the number of atoms of type $i$ that were added or removed to form the defect, and $\mu_i$ represents their corresponding chemical potentials, determined from their crystalline phase. $E_{VBM}$ is the energy of the valence band maximum (VBM) of the supercell containing the neutral defect and $E_{F}$ is the Fermi energy measured relative to the VBM. $E_{corr}$ is a correction term accounting for spurious interactions of charged defects under periodic boundary conditions, for which the self-consistent potential correction (SCPC) scheme implemented in VASP was used. \cite{Method_SCPC}

For band structure calculations and further analyses, $5\times 5$ supercells with hexagonal symmetry were employed. Since the SCPC scheme currently supports only orthorhombic cells, the cell symmetry was transformed from hexagonal to orthorhombic for the calculation of formation energies. $6\times 6$ in-plane supercells were utilized. In both cases, a c-axis lattice constant of $10\AA$ was chosen. The supercell sizes and the c-axis lattice constant were determined based on benchmark calculations (See Fig.~\ref{SI_Eformcaxis}).

\section{Results and Discussion}
\subsection{Formation energies}
\begin{table}[!htb]
    \centering
    \begin{tabular}{cc|cc|cc}
    Defect        &$E_{form}$ / eV&Defect         &$E_{form}$ / eV&Defect         &$E_{form}$ / eV \\
    \hline
    \vspace{0.05em} 
    -             &-              &\defect{Ce}{Mo}&1.16           &\defect{Er}{Mo}&1.06\\
    \defect{V}{S} &2.64           &\defect{V}{S}  &2.97           &\defect{V}{S}  &2.97\\
    \defect{V}{2S}&5.22           &\defect{V}{2S} &4.76           &\defect{V}{2S} &4.84\\
    \defect{V}{3S}&8.08           &\defect{V}{3S} &6.59           &\defect{V}{3S} &6.58\\
    \hline
    \end{tabular}       
    \caption{The calculated formation energies of \defect{MoS}{2} monolayer with defect compositions.}
    \label{tab:formationEnergies}
\end{table}

\begin{figure}[b]
    \centering
    \includegraphics[width=0.45\textwidth]{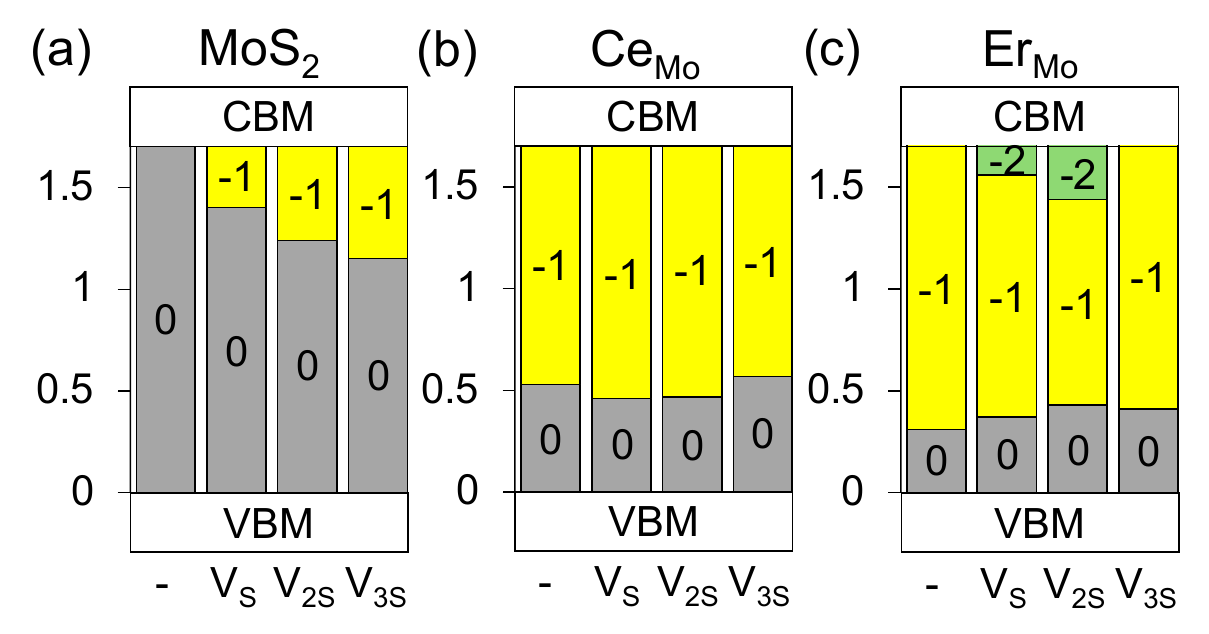}
    \caption{Stable charge states with respect to the Fermi energy ($E_F$) of the defect compositions within the valence band maximum (VBM) and conduction band minimum (CBM) in pristine \defect{MoS}{2} monolayer. (from left) \defect{MoS}{2} monolayer, \defect{V}{S}, \defect{V}{2S}, \defect{V}{3S}, \defect{Ce}{Mo}, \defect{Ce}{Mo}\defect{V}{S}, \defect{Ce}{Mo}\defect{V}{2S}, \defect{Ce}{Mo}\defect{V}{3S}, \defect{Er}{Mo}, \defect{Er}{Mo}\defect{V}{S}, \defect{Er}{Mo}\defect{V}{2S}, and \defect{Er}{Mo}\defect{V}{3S}.}
    \label{Fig_stable charge states within Eg}
\end{figure}

\begin{figure*}[t]
    \centering
    \includegraphics[width=0.9\textwidth]{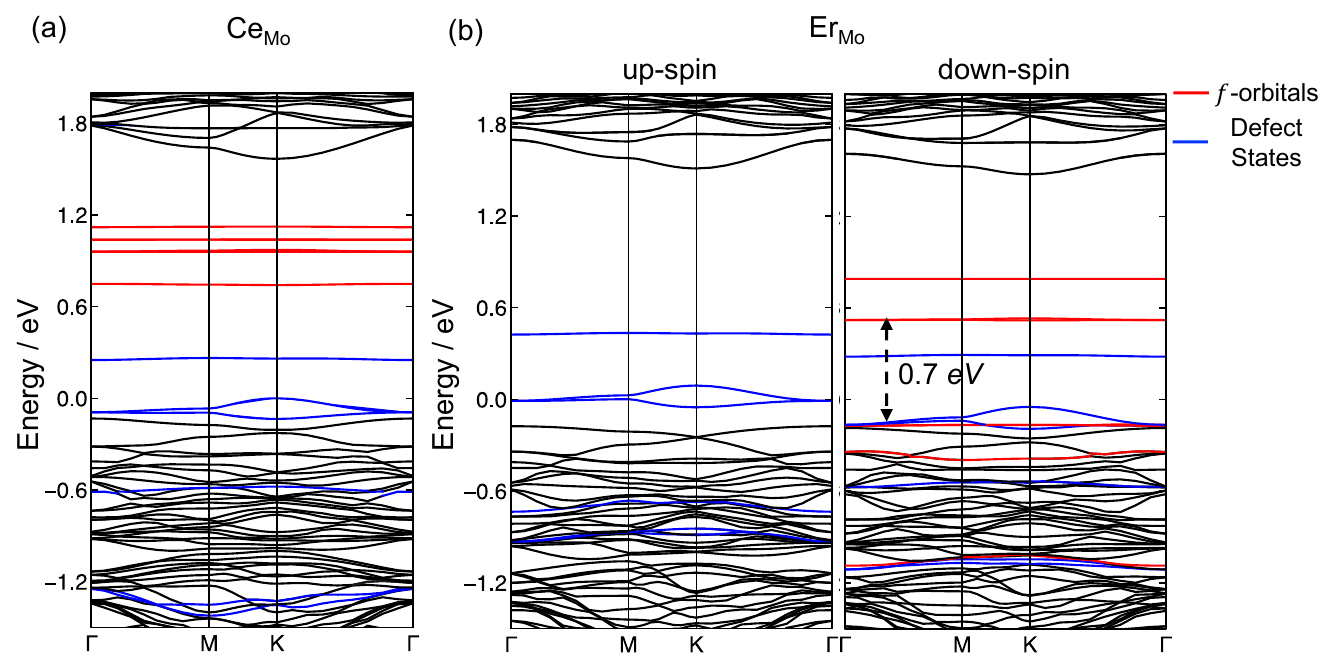}
    \caption{Band structures of (a) \defect{Ce}{Mo} and (b) \defect{Er}{Mo}. The red and blue bands show $f$-orbitals and defect states not originating from $f$-orbitals, respectively. The Fermi energy is set at $E_F = 0\ eV$.}
    \label{Fig_LnMoBS}
\end{figure*}

To investigate the effect of the larger ionic radii of the substituent atoms (Ln = Ce, Er) on the thermodynamic stability of \defect{MoS}{2}, we calculated the formation energies of neutral \defect{Ln}{Mo} with and without adjacent S vacancies (\defect{V}{S}, \defect{V}{2S}, and \defect{V}{3S}). For comparison, the formation energies of \defect{V}{S}, \defect{V}{2S}, and \defect{V}{3S} were also calculated (Table~\ref{tab:formationEnergies}). These values were determined using a chemical potential for S, $\mu_{S} = -4.1 eV$, approximated as the energy per atom of orthorhombic crystalline S and for Mo, $\mu_{Mo} = -10.8 eV$, estimated as the energy per atom of crystalline bcc-Mo. With these parameters, the formation energy of pristine \defect{MoS}{2} was calculated to be $E_{form} = E_{MoS_2} - \mu_{Mo} - 2 \mu_{S} = -2.81\ eV$, consistent with experimental and computational values reported in the literature ($-2.86\ eV$ and $-2.50\ eV$). \cite{VSMoS21_Eform_VMoPCHGD, VSMoS23_Eform, VSMoS25_Eform, VSMoS26_Eform} Similarly, the formation energy of \defect{V}{S} was calculated as $2.64\ eV$, also consistent with previous estimates ($2.35\ eV$ to $2.90\ eV$).

The formation energies of \defect{Ln}{Mo} without adjacent S vacancies were calculated to be $1.16\ eV$ for \defect{Ce}{Mo} and $1.06\ eV$ for \defect{Er}{Mo} (Table~\ref{tab:formationEnergies}). The size difference between Mo and Ln atoms becomes apparent when comparing the formation energies of \defect{MoS}{2} with identical numbers of S vacancies, both with and without \defect{Ln}{Mo} substitution.

For instance, the formation energy of \defect{Ln}{Mo}\defect{V}{S} ($2.97\ eV$) is $0.33\ eV$ higher than that of \defect{V}{S} ($2.64\ eV$), indicating the need for more space to accommodate the larger Ln ions. Moreover, as additional S vacancies are introduced, the formation energies in presence of \defect{Ln}{Mo} ($4.76\ eV$ to $6.59\ eV$) become lower than the cases without lanthanide substitution ($5.22\ eV$ to $8.08\ eV$). This indicates that neighboring S vacancies can mitigate the lattice strain induced by the larger lanthanide ions, thereby enhancing the thermodynamic stability of \defect{Ln}{Mo}.

These findings suggest that additional S vacancies may be required to accommodate the substitution of larger lanthanide ions, such as Ce or Er. This could be experimentally validated, as S vacancies are well known in \defect{MoS}{2} and can be controlled under experimental conditions. \cite{VSMoS21_Eform_VMoPCHGD, VSMoS22, VSMoS23_Eform, VSMoS24, VSMoS25_Eform, VSMoS26_Eform,VSMoS27}

Neutral defects model the isoelectronic substitution of Mo$^{4+}$ with Ln$^{4+}$. However, lanthanides typically exist as Ln$^{3+}$, corresponding to negatively charged defects. To explain the role of charge, we calculated formation energies for various charge states ($q$) as a function of the Fermi level ($E_F$) using Eq.~\ref{eq1}. At least two stable charge states were observed within the band gap of \defect{MoS}{2} for the considered defects (Fig.~\ref{Fig_stable charge states within Eg}). The band gap of \defect{MoS}{2} was calculated to be $1.7\ eV$ (Fig.~\ref{SI_Pristine BS}), consistent with other theoretical values and reflective of known DFT limitations. \cite{VSMoS27, MoS2Eg1, MoS2Eg2, MoS2Eg3} S vacancies, as reported in prior studies, exhibit n-type behavior due to the presence of defect states close to the CMB, stabilizing negative charge states near the conduction band edge ($E_F \sim 1.6\ eV$), as shown in Fig.~\ref{Fig_stable charge states within Eg}(a). \cite{VSMoS21_Eform_VMoPCHGD, VSMoS23_Eform, VSMoS25_Eform, VSMoS26_Eform}

However, for \defect{Ln}{Mo}, a negative charge state is already stabilized for $E_F \sim 0.5\ eV$, roughly independent of the presence of additional adjacent S vacancies [ Fig.~\ref{Fig_stable charge states within Eg}(b) and (c)]. This indicates that defect states introduced by \defect{Ln}{Mo} are located near the VBM and play a dominant role in charged state thermodynamics.

\subsection{Band Structures and Details of Defect States}
\begin{figure*}[t]
    \centering
    \includegraphics[width=0.9\textwidth]{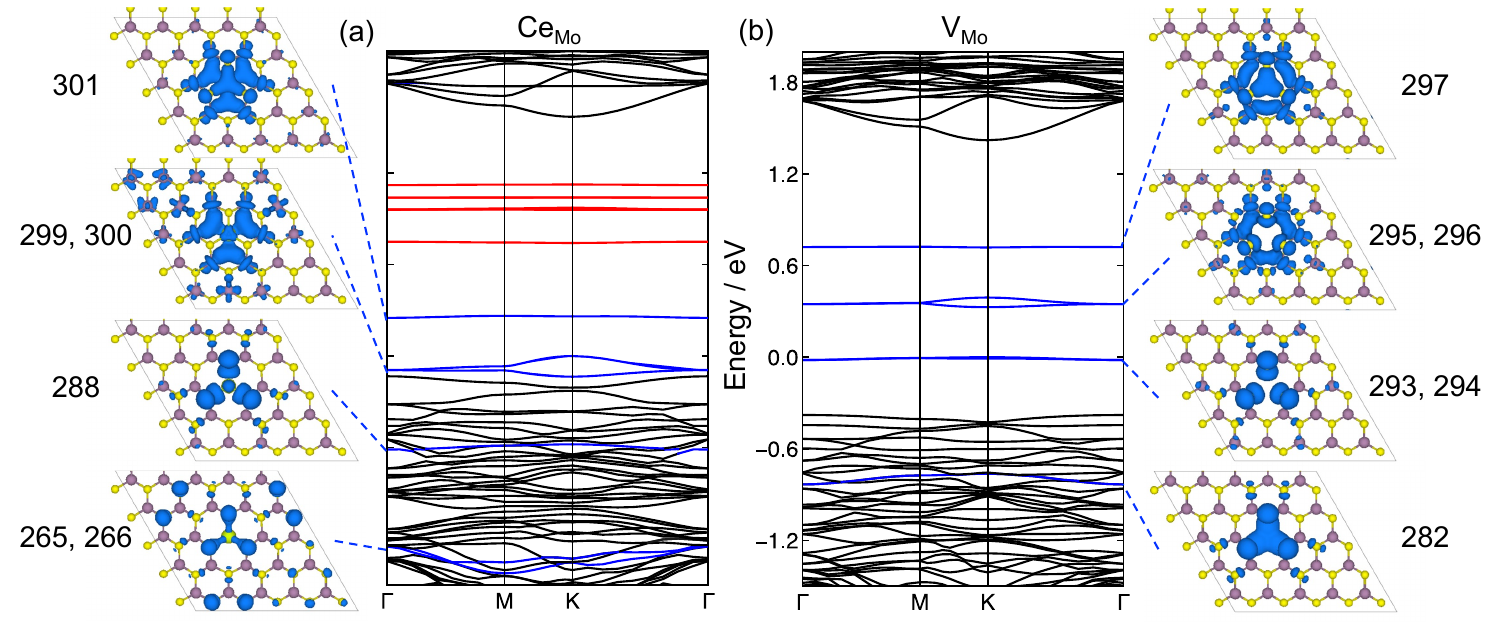}
    \caption{Band structures and charge distributions of defect states (blue bands) of (a) \defect{Ce}{Mo} and (b) \defect{V}{Mo}. The red and blue bands show $f$-orbitals and defect states not originating from $f$-orbitals, respectively. The Fermi energy is set at $E_F = 0\ eV$. The charge density isosurface value is set to 0.001.}
    \label{Fig_CeMoVMoPCHGD}
\end{figure*}

To comprehensively understand the impact of lanthanide ion doping on the electronic structure of \defect{MoS}{2}, we employed band structure projections for each orbital in conjunction with Partial Charged Density (PCHGD) calculations. These calculations provide detailed charge density distributions for each eigenstate, offering insights into the occupancy of atomic orbitals derived from individual atoms. This data enables an in-depth analysis of the orbital characteristics of each eigenstate, particularly those associated with defects. Given the highly localized nature of defects, their eigenstates are primarily influenced by local atomic orbitals. 

In contrast to most other lanthanides, Ce is known to exist in two electron configurations, Ce$^{3+}$ ([Xe]$4f^1$) and Ce$^{4+}$ ([Xe]). Thus, the neutral \defect{Ln}{Mo} corresponds to a formal Ce$^{4+}$ configuration in which all $f$-orbitals are unoccupied. These seven empty $f$-orbitals are expected to appear as localized states within the band gap of \defect{MoS}{2}. As shown in Fig.~\ref{Fig_LnMoBS}(a), the band structure confirms the presence of these defect states. The splitting of $f$-orbitals follows predictions from crystal field theory, as Ce is six-fold coordinated by S atoms in a prismatic structure with $D_{3h}$ point group symmetry [Fig.~\ref{SI_forbital}(a)].

However, in addition to the expected $f$-orbitals, additional localized defect states are observed near the VBM. These defect states, as discussed in greater detail below, do not originate from $f$-orbitals but instead exhibit $d-p$ character due to hybridization with dangling bonds from surrounding atoms. Moreover, in the band structure of negatively charged \defect{Ce}{Mo}, corresponding to a formal Ce$^{3+}$ configuration, an additional electron occupies one of these defect states rather than the $f$-orbitals. This result, shown in Fig.~\ref{SI_LnMo-}(a), indicates that the $f$-orbitals remain empty even for a negatively charged defect. This finding has significant implications for the use of \defect{Ce}{Mo}-type defects as SPE, as the desired $f-f$ transitions are predicted to be absent due to the filling of the additional defect states.

\begin{figure*}[t]
    \centering
    \includegraphics[width=0.9\textwidth]{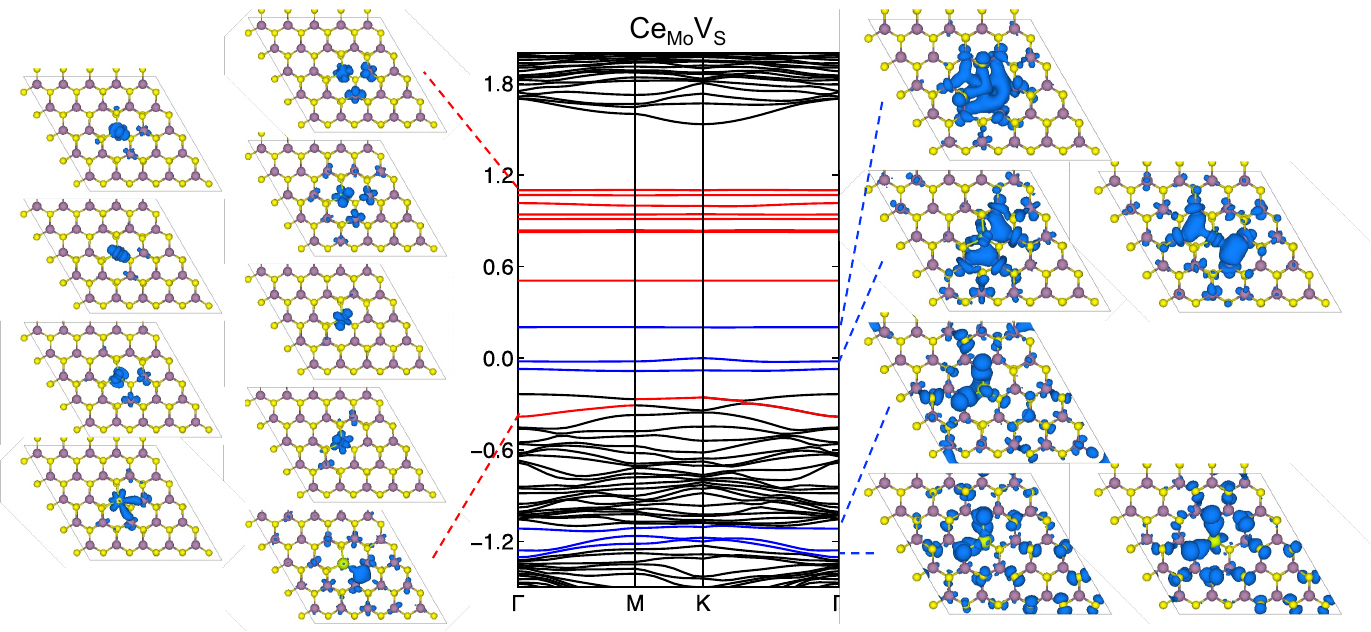}
    \caption{Band structures and charge distributions of defect states (red bands to left and blue bands to right) of \defect{Ce}{Mo}\defect{V}{S}. The red bands correspond to states originating from the hybridization of $f$-orbitals with \defect{V}{S} defect states. The blue bands correspond to the defect states of \defect{Ce}{Mo}. The Fermi energy is set at $E_F = 0\ eV$. The charge density isosurface value is set to 0.003 for red bands and 0.001 for blue bands respectively.}
    \label{Fig_CeMoVS}
\end{figure*}

Unlike Ce, Er generally exists in a formal $3+$ charge state, Er$^{3+}$ ([Xe]$4f^{11}$). The band structures for up-spin and down-spin states of \defect{Er}{Mo} are shown in Fig.~\ref{Fig_LnMoBS}(b). The differences between the two spin channels arise from a magnetic moment of $3\mu_b$. The splitting of $f$-orbitals is consistent with a local $D_{3h}$ symmetry [Fig..~\ref{SI_forbital}(b)]. For the down-spin channel, where not all seven $f$-orbitals are fully occupied, the energy difference between the highest occupied molecular orbital (HOMO) and the lowest unoccupied molecular orbital (LUMO) within the $f$-orbitals is approximately $0.7\ eV$. Including SOC, this energy difference is calculated to be about $0.9\ eV$. This value is consistent with the calculated value of $0.91\ eV$ for \defect{Er}{W}\defect{WS}{2}. \cite{LnTMDCSPE3_WS2}

The formal $3+$ charge state of Er (Er$^{3+}$) and its electronic configuration ([Xe]$4f^{11}$) remains unchanged in both the neutral and negatively charged state of \defect{Er}{Mo} [Fig.~\ref{Fig_LnMoBS}(b) and \ref{SI_LnMo-}(b)], consistent with previous observations for \defect{Er}{W}\defect{WS}{2}. \cite{LnTMDCSPE3_WS2} Similar to \defect{Ce}{Mo}, the $f$-orbitals of Er are not involved in electron redistribution upon a change in the charge state of the supercell due to the presence of additional defect states.

In addition to the defect state with $d$-$p$ character near the VBM, several other defect states are observed in the band structures of \defect{Ce}{Mo} [Fig.~\ref{Fig_LnMoBS}(a)] and \defect{Er}{Mo} [Fig.~\ref{Fig_LnMoBS}(b)]. These include three states within the valence bands and three within the band gap. Charge distribution analysis reveals that the features and ordering of these states in \defect{Ce}{Mo} [Fig.~\ref{Fig_CeMoVMoPCHGD}(a)] are similar to those in \defect{Er}{Mo} (\ref{SI_ErMoPCHGD}).

The charge distribution of defect states in \defect{Ce}{Mo}, shown in Fig.~\ref{Fig_CeMoVMoPCHGD}(a), reveals a common structural feature: the surrounding atoms form a ring-like arrangement centered around the Ce atom. Defect states within the valence bands (e.g., 265, 266, and 288) exhibit $p$-character originating from S atoms, while those within the band gap (e.g., 299, 300, and 301) exhibit $d$-$p$ character.

Notably, these defect states closely resemble those of \defect{V}{Mo}, as reported by \citet{VSMoS21_Eform_VMoPCHGD}. In \defect{V}{Mo}, six dangling bond states were identified: three arising from the hybridization of $p$-orbitals of adjacent S atoms and three from the hybridization of Mo-$d$ and S-$p$ orbitals. Our calculations reproduce these states [Fig.~\ref{Fig_CeMoVMoPCHGD}(b)].

Defect states generally arise from either disruption in chemical bonding or nonbonding orbital interactions, resulting in localized states and flat bands. We propose that the similarity between defect states in \defect{Ln}{Mo} and \defect{V}{Mo} stems from the mismatch in $d$-orbital electron configurations between the substituent and the host material. For Mo$^{4+}$, the electron configuration is [Kr]$4d^2$, whereas Ce and Er possess a $5d^0$ configuration. This mismatch prevents effective hybridization between the host material's orbitals and the substituent, leading to discontinuities in covalent bonds and the formation of dangling bonds.

This phenomenon is not exclusive to Ce and Er but extends to other ions with a $d^0$ configuration, such as \defect{La}{Mo} (La$^{3+}$ [Xe]), \defect{Lu}{Mo} (Lu$^{3+}$ [Xe]$4f^{14}$), and \defect{Zr}{Mo} (Zr$^{4+}$ [Kr]) [Fig.~\ref{SI_LaMoWMo}(a), (b), and (c)]. Conversely, no defect states are observed in the isoelectronic \defect{W}{Mo} (W$^{4+}$ [Xe]$d^2$) [Fig.\ref{SI_LaMoWMo}(d)].

The qualitative features of these defect states remain consistent in the presence of additional defects. For example, similar states are observed in the band structures of \defect{Ce}{Mo}\defect{V}{S} (Fig.~\ref{Fig_CeMoVS}). While S vacancies alter symmetry and electron count, one can still identify charge distributions that resemble those of \defect{Ce}{Mo}.

These results suggest that an electron configuration mismatch between the substituent and the host material may introduce additional, potentially undesirable, defect states that interfere with the optical properties of point defects. For instance, if a SPE based on Ce$^{3+}$ with the electron configuration of [Xe]$4f^1$ is desired, a host material without orbital occupation mismatch, such as TMDCs with an electron configuration of $d^0$, would likely be preferable. 

This observation is highly relevant for embedding methods, that describe defect states using a higher level of theory, embedded into the host material that is described by a lower level of theory. 
For example, one would expect only states in the bandgap derived from f-orbitals, and would describe aim to include theses states explicitly. However, as we show, states that derive mainly from the host material also appear in the gap, and are vital to understand the electronic and structure properties of the defect. Unlike the f-states, these states are more delocalized and thus likely interact more with the host materials. It remains to be shown if embedding methods can faithfully describe them, as the active (embedded) space needs to incorporate more atoms than just those close to the defect.


\section{Conclusion}
In this study, we employ density functional theory (DFT) calculations to investigate the structural and electronic properties of lanthanide-doped \defect{MoS}{2} monolayers, focusing on Ce and Er substitution at the Mo site. Our results show that the substituting larger lanthanide ions destabilize pristine \defect{MoS}{2}, as indicated by higher formation energies. However, introducing additional S vacancies significantly mitigates this destabilization by relieving lattice strain, highlighting the critical role of defect engineering in realizing \defect{Ln}{Mo} in experiments. Charge state analysis reveals that lanthanide dopants introduce defect states near the valence band maximum, supporting stable negative charge states near the conduction band edge. 

Electronic structure calculations indicate that Ce consistently adopts a +4 charge state, whereas Er remains in a +3 charge state, regardless of the supercell charge. Due to the absence of $f$-electrons in Ce$^{4+}$, no significant $f$-$f$ electron transitions are expected. In contrast, Er$^{3+}$ exhibits an $f$-$f$ electron transition at approximately $0.7\ eV$, suggesting its potential for quantum telecommunication applications as a SPEs.

Band structure projections and projected charge density (PCHGD) calculations reveal that, in addition to $f$-orbital-related states, defect states associated with the host material emerge near the Fermi level in \defect{Ln}{Mo}, contributing to a range of electronic transitions. This behavior is attributed to the mismatch between the outermost $d$-orbital electron configurations of Mo$^{4+}$ ($d^2$) and the Ln ions ($d^0$), a trend observed in other transition-metal-doped TMDCs with varying $d$-orbital configurations.

Furthermore, these defect states exhibit qualitative similarities across different supercell symmetries and electron counts, despite the presence of additional neighboring defects. This finding underscores the importance of considering both the electronic configuration of the dopant and its interaction with the host lattice when analyzing band structure phenomena.

Overall, these results demonstrate that lanthanide-doped \defect{MoS}{2} presents a viable platform for quantum telecommunications applications, provided that defect configurations are carefully engineered. This study lays the groundwork for future experimental and computational efforts to optimize TMDCs-based SPEs for integration into quantum technologies.

\bibliography{ref}

\begin{thebibliography}{53}%
\makeatletter
\providecommand \@ifxundefined [1]{%
 \@ifx{#1\undefined}
}%
\providecommand \@ifnum [1]{%
 \ifnum #1\expandafter \@firstoftwo
 \else \expandafter \@secondoftwo
 \fi
}%
\providecommand \@ifx [1]{%
 \ifx #1\expandafter \@firstoftwo
 \else \expandafter \@secondoftwo
 \fi
}%
\providecommand \natexlab [1]{#1}%
\providecommand \enquote  [1]{``#1''}%
\providecommand \bibnamefont  [1]{#1}%
\providecommand \bibfnamefont [1]{#1}%
\providecommand \citenamefont [1]{#1}%
\providecommand \href@noop [0]{\@secondoftwo}%
\providecommand \href [0]{\begingroup \@sanitize@url \@href}%
\providecommand \@href[1]{\@@startlink{#1}\@@href}%
\providecommand \@@href[1]{\endgroup#1\@@endlink}%
\providecommand \@sanitize@url [0]{\catcode `\\12\catcode `\$12\catcode `\&12\catcode `\#12\catcode `\^12\catcode `\_12\catcode `\%12\relax}%
\providecommand \@@startlink[1]{}%
\providecommand \@@endlink[0]{}%
\providecommand \url  [0]{\begingroup\@sanitize@url \@url }%
\providecommand \@url [1]{\endgroup\@href {#1}{\urlprefix }}%
\providecommand \urlprefix  [0]{URL }%
\providecommand \Eprint [0]{\href }%
\providecommand \doibase [0]{https://doi.org/}%
\providecommand \selectlanguage [0]{\@gobble}%
\providecommand \bibinfo  [0]{\@secondoftwo}%
\providecommand \bibfield  [0]{\@secondoftwo}%
\providecommand \translation [1]{[#1]}%
\providecommand \BibitemOpen [0]{}%
\providecommand \bibitemStop [0]{}%
\providecommand \bibitemNoStop [0]{.\EOS\space}%
\providecommand \EOS [0]{\spacefactor3000\relax}%
\providecommand \BibitemShut  [1]{\csname bibitem#1\endcsname}%
\let\auto@bib@innerbib\@empty
\bibitem [{\citenamefont {Kok}\ \emph {et~al.}(2007)\citenamefont {Kok}, \citenamefont {Munro}, \citenamefont {Nemoto}, \citenamefont {Ralph}, \citenamefont {Dowling},\ and\ \citenamefont {Milburn}}]{SPEforquantum1}%
  \BibitemOpen
  \bibfield  {author} {\bibinfo {author} {\bibfnamefont {P.}~\bibnamefont {Kok}}, \bibinfo {author} {\bibfnamefont {W.~J.}\ \bibnamefont {Munro}}, \bibinfo {author} {\bibfnamefont {K.}~\bibnamefont {Nemoto}}, \bibinfo {author} {\bibfnamefont {T.~C.}\ \bibnamefont {Ralph}}, \bibinfo {author} {\bibfnamefont {J.~P.}\ \bibnamefont {Dowling}},\ and\ \bibinfo {author} {\bibfnamefont {G.~J.}\ \bibnamefont {Milburn}},\ }\bibfield  {title} {\bibinfo {title} {Linear optical quantum computing with photonic qubits},\ }\href {https://doi.org/10.1103/RevModPhys.79.135} {\bibfield  {journal} {\bibinfo  {journal} {Rev. Mod. Phys.}\ }\textbf {\bibinfo {volume} {79}},\ \bibinfo {pages} {135} (\bibinfo {year} {2007})}\BibitemShut {NoStop}%
\bibitem [{\citenamefont {Gimeno-Segovia}\ \emph {et~al.}(2015)\citenamefont {Gimeno-Segovia}, \citenamefont {Shadbolt}, \citenamefont {Browne},\ and\ \citenamefont {Rudolph}}]{SPEforquantum2}%
  \BibitemOpen
  \bibfield  {author} {\bibinfo {author} {\bibfnamefont {M.}~\bibnamefont {Gimeno-Segovia}}, \bibinfo {author} {\bibfnamefont {P.}~\bibnamefont {Shadbolt}}, \bibinfo {author} {\bibfnamefont {D.~E.}\ \bibnamefont {Browne}},\ and\ \bibinfo {author} {\bibfnamefont {T.}~\bibnamefont {Rudolph}},\ }\bibfield  {title} {\bibinfo {title} {From three-photon greenberger-horne-zeilinger states to ballistic universal quantum computation},\ }\href {https://doi.org/10.1103/PhysRevLett.115.020502} {\bibfield  {journal} {\bibinfo  {journal} {Phys. Rev. Lett.}\ }\textbf {\bibinfo {volume} {115}},\ \bibinfo {pages} {020502} (\bibinfo {year} {2015})}\BibitemShut {NoStop}%
\bibitem [{\citenamefont {Aspuru-Guzik}\ and\ \citenamefont {Walther}(2012)}]{SPEforquantum3}%
  \BibitemOpen
  \bibfield  {author} {\bibinfo {author} {\bibfnamefont {A.}~\bibnamefont {Aspuru-Guzik}}\ and\ \bibinfo {author} {\bibfnamefont {P.}~\bibnamefont {Walther}},\ }\bibfield  {title} {\bibinfo {title} {Photonic quantum simulators},\ }\href {https://doi.org/10.1038/nphys2253} {\bibfield  {journal} {\bibinfo  {journal} {Nature Physics}\ }\textbf {\bibinfo {volume} {8}} (\bibinfo {year} {2012})}\BibitemShut {NoStop}%
\bibitem [{\citenamefont {Giovannetti}\ \emph {et~al.}(2011)\citenamefont {Giovannetti}, \citenamefont {Lloyd},\ and\ \citenamefont {Maccone}}]{SPEforquantum4}%
  \BibitemOpen
  \bibfield  {author} {\bibinfo {author} {\bibfnamefont {V.}~\bibnamefont {Giovannetti}}, \bibinfo {author} {\bibfnamefont {S.}~\bibnamefont {Lloyd}},\ and\ \bibinfo {author} {\bibfnamefont {L.}~\bibnamefont {Maccone}},\ }\bibfield  {title} {\bibinfo {title} {Advances in quantum metrology},\ }\href {https://doi.org/10.1038/nphoton.2011.35} {\bibfield  {journal} {\bibinfo  {journal} {Phys. Rev. Lett.}\ }\textbf {\bibinfo {volume} {96}} (\bibinfo {year} {2011})}\BibitemShut {NoStop}%
\bibitem [{\citenamefont {Scarani}\ \emph {et~al.}(2009)\citenamefont {Scarani}, \citenamefont {Bechmann-Pasquinucci}, \citenamefont {Cerf}, \citenamefont {Du\ifmmode~\check{s}\else \v{s}\fi{}ek}, \citenamefont {L\"utkenhaus},\ and\ \citenamefont {Peev}}]{SPEforquantum5}%
  \BibitemOpen
  \bibfield  {author} {\bibinfo {author} {\bibfnamefont {V.}~\bibnamefont {Scarani}}, \bibinfo {author} {\bibfnamefont {H.}~\bibnamefont {Bechmann-Pasquinucci}}, \bibinfo {author} {\bibfnamefont {N.~J.}\ \bibnamefont {Cerf}}, \bibinfo {author} {\bibfnamefont {M.}~\bibnamefont {Du\ifmmode~\check{s}\else \v{s}\fi{}ek}}, \bibinfo {author} {\bibfnamefont {N.}~\bibnamefont {L\"utkenhaus}},\ and\ \bibinfo {author} {\bibfnamefont {M.}~\bibnamefont {Peev}},\ }\bibfield  {title} {\bibinfo {title} {The security of practical quantum key distribution},\ }\href {https://doi.org/10.1103/RevModPhys.81.1301} {\bibfield  {journal} {\bibinfo  {journal} {Rev. Mod. Phys.}\ }\textbf {\bibinfo {volume} {81}},\ \bibinfo {pages} {1301} (\bibinfo {year} {2009})}\BibitemShut {NoStop}%
\bibitem [{\citenamefont {Lo}\ \emph {et~al.}(2014)\citenamefont {Lo}, \citenamefont {Curty},\ and\ \citenamefont {Tamaki}}]{SPEforquantum6}%
  \BibitemOpen
  \bibfield  {author} {\bibinfo {author} {\bibfnamefont {H.-K.}\ \bibnamefont {Lo}}, \bibinfo {author} {\bibfnamefont {M.}~\bibnamefont {Curty}},\ and\ \bibinfo {author} {\bibfnamefont {K.}~\bibnamefont {Tamaki}},\ }\bibfield  {title} {\bibinfo {title} {Secure quantum key distribution},\ }\href {https://api.semanticscholar.org/CorpusID:119105614} {\bibfield  {journal} {\bibinfo  {journal} {Nature Photonics}\ }\textbf {\bibinfo {volume} {8}},\ \bibinfo {pages} {595 } (\bibinfo {year} {2014})}\BibitemShut {NoStop}%
\bibitem [{\citenamefont {Koehl}\ \emph {et~al.}(2015)\citenamefont {Koehl}, \citenamefont {Seo}, \citenamefont {Galli},\ and\ \citenamefont {Awschalom}}]{solidSPE1}%
  \BibitemOpen
  \bibfield  {author} {\bibinfo {author} {\bibfnamefont {W.~F.}\ \bibnamefont {Koehl}}, \bibinfo {author} {\bibfnamefont {H.}~\bibnamefont {Seo}}, \bibinfo {author} {\bibfnamefont {G.}~\bibnamefont {Galli}},\ and\ \bibinfo {author} {\bibfnamefont {D.~D.}\ \bibnamefont {Awschalom}},\ }\bibfield  {title} {\bibinfo {title} {Designing defect spins for wafer-scale quantum technologies},\ }\href@noop {} {\bibfield  {journal} {\bibinfo  {journal} {MRS Bulletin}\ }\textbf {\bibinfo {volume} {40}},\ \bibinfo {pages} {1146} (\bibinfo {year} {2015})}\BibitemShut {NoStop}%
\bibitem [{\citenamefont {Loredo}\ \emph {et~al.}(2016)\citenamefont {Loredo}, \citenamefont {Zakaria}, \citenamefont {Somaschi}, \citenamefont {Anton}, \citenamefont {de~Santis}, \citenamefont {Giesz}, \citenamefont {Grange}, \citenamefont {Broome}, \citenamefont {Gazzano}, \citenamefont {Coppola}, \citenamefont {Sagnes}, \citenamefont {Lemaitre}, \citenamefont {Auffeves}, \citenamefont {Senellart}, \citenamefont {Almeida},\ and\ \citenamefont {White}}]{solidSPE2}%
  \BibitemOpen
  \bibfield  {author} {\bibinfo {author} {\bibfnamefont {J.~C.}\ \bibnamefont {Loredo}}, \bibinfo {author} {\bibfnamefont {N.~A.}\ \bibnamefont {Zakaria}}, \bibinfo {author} {\bibfnamefont {N.}~\bibnamefont {Somaschi}}, \bibinfo {author} {\bibfnamefont {C.}~\bibnamefont {Anton}}, \bibinfo {author} {\bibfnamefont {L.}~\bibnamefont {de~Santis}}, \bibinfo {author} {\bibfnamefont {V.}~\bibnamefont {Giesz}}, \bibinfo {author} {\bibfnamefont {T.}~\bibnamefont {Grange}}, \bibinfo {author} {\bibfnamefont {M.~A.}\ \bibnamefont {Broome}}, \bibinfo {author} {\bibfnamefont {O.}~\bibnamefont {Gazzano}}, \bibinfo {author} {\bibfnamefont {G.}~\bibnamefont {Coppola}}, \bibinfo {author} {\bibfnamefont {I.}~\bibnamefont {Sagnes}}, \bibinfo {author} {\bibfnamefont {A.}~\bibnamefont {Lemaitre}}, \bibinfo {author} {\bibfnamefont {A.}~\bibnamefont {Auffeves}}, \bibinfo {author} {\bibfnamefont {P.}~\bibnamefont {Senellart}}, \bibinfo {author} {\bibfnamefont {M.~P.}\ \bibnamefont {Almeida}},\ and\ \bibinfo {author} {\bibfnamefont
  {A.~G.}\ \bibnamefont {White}},\ }\bibfield  {title} {\bibinfo {title} {Scalable performance in solid-state single-photon sources},\ }\href {https://doi.org/10.1364/OPTICA.3.000433} {\bibfield  {journal} {\bibinfo  {journal} {Optica}\ }\textbf {\bibinfo {volume} {3}},\ \bibinfo {pages} {433} (\bibinfo {year} {2016})}\BibitemShut {NoStop}%
\bibitem [{\citenamefont {Somaschi}\ \emph {et~al.}(2015)\citenamefont {Somaschi}, \citenamefont {Giesz}, \citenamefont {De~Santis}, \citenamefont {Loredo}, \citenamefont {Almeida}, \citenamefont {Hornecker}, \citenamefont {Portalupi}, \citenamefont {Grange}, \citenamefont {Antón}, \citenamefont {Demory}, \citenamefont {Gomez~Carbonell}, \citenamefont {Sagnes}, \citenamefont {Lanzillotti~Kimura}, \citenamefont {Lemaître}, \citenamefont {Auffèves}, \citenamefont {White}, \citenamefont {Lanco},\ and\ \citenamefont {Senellart}}]{solidSPE3}%
  \BibitemOpen
  \bibfield  {author} {\bibinfo {author} {\bibfnamefont {N.}~\bibnamefont {Somaschi}}, \bibinfo {author} {\bibfnamefont {V.}~\bibnamefont {Giesz}}, \bibinfo {author} {\bibfnamefont {L.}~\bibnamefont {De~Santis}}, \bibinfo {author} {\bibfnamefont {J.}~\bibnamefont {Loredo}}, \bibinfo {author} {\bibfnamefont {M.}~\bibnamefont {Almeida}}, \bibinfo {author} {\bibfnamefont {G.}~\bibnamefont {Hornecker}}, \bibinfo {author} {\bibfnamefont {S.}~\bibnamefont {Portalupi}}, \bibinfo {author} {\bibfnamefont {T.}~\bibnamefont {Grange}}, \bibinfo {author} {\bibfnamefont {C.}~\bibnamefont {Antón}}, \bibinfo {author} {\bibfnamefont {J.}~\bibnamefont {Demory}}, \bibinfo {author} {\bibfnamefont {C.}~\bibnamefont {Gomez~Carbonell}}, \bibinfo {author} {\bibfnamefont {I.}~\bibnamefont {Sagnes}}, \bibinfo {author} {\bibfnamefont {D.}~\bibnamefont {Lanzillotti~Kimura}}, \bibinfo {author} {\bibfnamefont {A.}~\bibnamefont {Lemaître}}, \bibinfo {author} {\bibfnamefont {A.}~\bibnamefont {Auffèves}}, \bibinfo {author} {\bibfnamefont
  {A.}~\bibnamefont {White}}, \bibinfo {author} {\bibfnamefont {L.}~\bibnamefont {Lanco}},\ and\ \bibinfo {author} {\bibfnamefont {P.}~\bibnamefont {Senellart}},\ }\bibfield  {title} {\bibinfo {title} {Near optimal single photon sources in the solid state},\ }\href {https://doi.org/10.1038/nphoton.2016.23} {\bibfield  {journal} {\bibinfo  {journal} {Nature Photonics}\ }\textbf {\bibinfo {volume} {10}} (\bibinfo {year} {2015})}\BibitemShut {NoStop}%
\bibitem [{\citenamefont {Ding}\ \emph {et~al.}(2016)\citenamefont {Ding}, \citenamefont {He}, \citenamefont {Duan}, \citenamefont {Gregersen}, \citenamefont {Chen}, \citenamefont {Unsleber}, \citenamefont {Maier}, \citenamefont {Schneider}, \citenamefont {Kamp}, \citenamefont {H\"ofling}, \citenamefont {Lu},\ and\ \citenamefont {Pan}}]{solidSPE4}%
  \BibitemOpen
  \bibfield  {author} {\bibinfo {author} {\bibfnamefont {X.}~\bibnamefont {Ding}}, \bibinfo {author} {\bibfnamefont {Y.}~\bibnamefont {He}}, \bibinfo {author} {\bibfnamefont {Z.-C.}\ \bibnamefont {Duan}}, \bibinfo {author} {\bibfnamefont {N.}~\bibnamefont {Gregersen}}, \bibinfo {author} {\bibfnamefont {M.-C.}\ \bibnamefont {Chen}}, \bibinfo {author} {\bibfnamefont {S.}~\bibnamefont {Unsleber}}, \bibinfo {author} {\bibfnamefont {S.}~\bibnamefont {Maier}}, \bibinfo {author} {\bibfnamefont {C.}~\bibnamefont {Schneider}}, \bibinfo {author} {\bibfnamefont {M.}~\bibnamefont {Kamp}}, \bibinfo {author} {\bibfnamefont {S.}~\bibnamefont {H\"ofling}}, \bibinfo {author} {\bibfnamefont {C.-Y.}\ \bibnamefont {Lu}},\ and\ \bibinfo {author} {\bibfnamefont {J.-W.}\ \bibnamefont {Pan}},\ }\bibfield  {title} {\bibinfo {title} {On-demand single photons with high extraction efficiency and near-unity indistinguishability from a resonantly driven quantum dot in a micropillar},\ }\href
  {https://doi.org/10.1103/PhysRevLett.116.020401} {\bibfield  {journal} {\bibinfo  {journal} {Phys. Rev. Lett.}\ }\textbf {\bibinfo {volume} {116}},\ \bibinfo {pages} {020401} (\bibinfo {year} {2016})}\BibitemShut {NoStop}%
\bibitem [{\citenamefont {Kane}(1998)}]{pointdefectSPE1}%
  \BibitemOpen
  \bibfield  {author} {\bibinfo {author} {\bibfnamefont {B.~E.}\ \bibnamefont {Kane}},\ }\bibfield  {title} {\bibinfo {title} {A silicon-based nuclear spin quantum computer},\ }\href {https://api.semanticscholar.org/CorpusID:8470520} {\bibfield  {journal} {\bibinfo  {journal} {Nature}\ }\textbf {\bibinfo {volume} {393}},\ \bibinfo {pages} {133} (\bibinfo {year} {1998})}\BibitemShut {NoStop}%
\bibitem [{\citenamefont {Weber}\ \emph {et~al.}(2010)\citenamefont {Weber}, \citenamefont {Koehl}, \citenamefont {Varley}, \citenamefont {Janotti}, \citenamefont {Buckley}, \citenamefont {Walle},\ and\ \citenamefont {Awschalom}}]{pointdefectSPE2}%
  \BibitemOpen
  \bibfield  {author} {\bibinfo {author} {\bibfnamefont {J.}~\bibnamefont {Weber}}, \bibinfo {author} {\bibfnamefont {W.}~\bibnamefont {Koehl}}, \bibinfo {author} {\bibfnamefont {J.}~\bibnamefont {Varley}}, \bibinfo {author} {\bibfnamefont {A.}~\bibnamefont {Janotti}}, \bibinfo {author} {\bibfnamefont {B.}~\bibnamefont {Buckley}}, \bibinfo {author} {\bibfnamefont {C.}~\bibnamefont {Walle}},\ and\ \bibinfo {author} {\bibfnamefont {D.}~\bibnamefont {Awschalom}},\ }\bibfield  {title} {\bibinfo {title} {Quantum computing with defects},\ }\href {https://doi.org/10.1073/pnas.1003052107} {\bibfield  {journal} {\bibinfo  {journal} {Proceedings of the National Academy of Sciences of the United States of America}\ }\textbf {\bibinfo {volume} {107}},\ \bibinfo {pages} {8513} (\bibinfo {year} {2010})}\BibitemShut {NoStop}%
\bibitem [{\citenamefont {Aharonovich}\ \emph {et~al.}(2011)\citenamefont {Aharonovich}, \citenamefont {Castelletto}, \citenamefont {Simpson}, \citenamefont {Su}, \citenamefont {Greentree},\ and\ \citenamefont {Prawer}}]{pointdefectSPE3}%
  \BibitemOpen
  \bibfield  {author} {\bibinfo {author} {\bibfnamefont {I.}~\bibnamefont {Aharonovich}}, \bibinfo {author} {\bibfnamefont {S.}~\bibnamefont {Castelletto}}, \bibinfo {author} {\bibfnamefont {D.~A.}\ \bibnamefont {Simpson}}, \bibinfo {author} {\bibfnamefont {C.-H.}\ \bibnamefont {Su}}, \bibinfo {author} {\bibfnamefont {A.~D.}\ \bibnamefont {Greentree}},\ and\ \bibinfo {author} {\bibfnamefont {S.}~\bibnamefont {Prawer}},\ }\bibfield  {title} {\bibinfo {title} {Diamond-based single-photon emitters},\ }\href {https://doi.org/10.1088/0034-4885/74/7/076501} {\bibfield  {journal} {\bibinfo  {journal} {Reports on Progress in Physics}\ }\textbf {\bibinfo {volume} {74}},\ \bibinfo {pages} {076501} (\bibinfo {year} {2011})}\BibitemShut {NoStop}%
\bibitem [{\citenamefont {Pla}\ \emph {et~al.}(2012)\citenamefont {Pla}, \citenamefont {Tan}, \citenamefont {Dehollain}, \citenamefont {Lim}, \citenamefont {Morton}, \citenamefont {Jamieson}, \citenamefont {Dzurak},\ and\ \citenamefont {Morello}}]{pointdefectSPE4}%
  \BibitemOpen
  \bibfield  {author} {\bibinfo {author} {\bibfnamefont {J.}~\bibnamefont {Pla}}, \bibinfo {author} {\bibfnamefont {K.}~\bibnamefont {Tan}}, \bibinfo {author} {\bibfnamefont {J.~P.}\ \bibnamefont {Dehollain}}, \bibinfo {author} {\bibfnamefont {W.}~\bibnamefont {Lim}}, \bibinfo {author} {\bibfnamefont {J.}~\bibnamefont {Morton}}, \bibinfo {author} {\bibfnamefont {D.}~\bibnamefont {Jamieson}}, \bibinfo {author} {\bibfnamefont {A.}~\bibnamefont {Dzurak}},\ and\ \bibinfo {author} {\bibfnamefont {A.}~\bibnamefont {Morello}},\ }\bibfield  {title} {\bibinfo {title} {A single-atom electron spin qubit in silicon},\ }\href {https://doi.org/10.1038/nature11449} {\bibfield  {journal} {\bibinfo  {journal} {Nature}\ }\textbf {\bibinfo {volume} {489}},\ \bibinfo {pages} {541} (\bibinfo {year} {2012})}\BibitemShut {NoStop}%
\bibitem [{\citenamefont {Wu}\ \emph {et~al.}(2019{\natexlab{a}})\citenamefont {Wu}, \citenamefont {Wang}, \citenamefont {Qin}, \citenamefont {Rong},\ and\ \citenamefont {Du}}]{pointdefectSPE5}%
  \BibitemOpen
  \bibfield  {author} {\bibinfo {author} {\bibfnamefont {Y.}~\bibnamefont {Wu}}, \bibinfo {author} {\bibfnamefont {Y.}~\bibnamefont {Wang}}, \bibinfo {author} {\bibfnamefont {X.}~\bibnamefont {Qin}}, \bibinfo {author} {\bibfnamefont {X.}~\bibnamefont {Rong}},\ and\ \bibinfo {author} {\bibfnamefont {J.}~\bibnamefont {Du}},\ }\bibfield  {title} {\bibinfo {title} {A programmable two-qubit solid-state quantum processor under ambient conditions},\ }\href {https://doi.org/10.1038/s41534-019-0129-z} {\bibfield  {journal} {\bibinfo  {journal} {npj Quantum Information}\ }\textbf {\bibinfo {volume} {5}},\ \bibinfo {pages} {9} (\bibinfo {year} {2019}{\natexlab{a}})}\BibitemShut {NoStop}%
\bibitem [{\citenamefont {Muechler}\ \emph {et~al.}(2020)\citenamefont {Muechler}, \citenamefont {Hu}, \citenamefont {Lin}, \citenamefont {Yang},\ and\ \citenamefont {Car}}]{pointdefectSPE6}%
  \BibitemOpen
  \bibfield  {author} {\bibinfo {author} {\bibfnamefont {L.}~\bibnamefont {Muechler}}, \bibinfo {author} {\bibfnamefont {W.}~\bibnamefont {Hu}}, \bibinfo {author} {\bibfnamefont {L.}~\bibnamefont {Lin}}, \bibinfo {author} {\bibfnamefont {C.}~\bibnamefont {Yang}},\ and\ \bibinfo {author} {\bibfnamefont {R.}~\bibnamefont {Car}},\ }\bibfield  {title} {\bibinfo {title} {Influence of point defects on the electronic and topological properties of monolayer ${\mathrm{wte}}_{2}$},\ }\href {https://doi.org/10.1103/PhysRevB.102.041103} {\bibfield  {journal} {\bibinfo  {journal} {Phys. Rev. B}\ }\textbf {\bibinfo {volume} {102}},\ \bibinfo {pages} {041103} (\bibinfo {year} {2020})}\BibitemShut {NoStop}%
\bibitem [{\citenamefont {Aharonovich}\ and\ \citenamefont {Neu}(2014)}]{NV-SPE1}%
  \BibitemOpen
  \bibfield  {author} {\bibinfo {author} {\bibfnamefont {I.}~\bibnamefont {Aharonovich}}\ and\ \bibinfo {author} {\bibfnamefont {E.}~\bibnamefont {Neu}},\ }\bibfield  {title} {\bibinfo {title} {Diamond nanophotonics},\ }\href {https://doi.org/https://doi.org/10.1002/adom.201400189} {\bibfield  {journal} {\bibinfo  {journal} {Advanced Optical Materials}\ }\textbf {\bibinfo {volume} {2}},\ \bibinfo {pages} {911} (\bibinfo {year} {2014})},\ \Eprint {https://arxiv.org/abs/https://onlinelibrary.wiley.com/doi/pdf/10.1002/adom.201400189} {https://onlinelibrary.wiley.com/doi/pdf/10.1002/adom.201400189} \BibitemShut {NoStop}%
\bibitem [{\citenamefont {Alkauskas}\ \emph {et~al.}(2014)\citenamefont {Alkauskas}, \citenamefont {Buckley}, \citenamefont {Awschalom},\ and\ \citenamefont {de~Walle}}]{NV-SPE2}%
  \BibitemOpen
  \bibfield  {author} {\bibinfo {author} {\bibfnamefont {A.}~\bibnamefont {Alkauskas}}, \bibinfo {author} {\bibfnamefont {B.~B.}\ \bibnamefont {Buckley}}, \bibinfo {author} {\bibfnamefont {D.~D.}\ \bibnamefont {Awschalom}},\ and\ \bibinfo {author} {\bibfnamefont {C.~G.~V.}\ \bibnamefont {de~Walle}},\ }\bibfield  {title} {\bibinfo {title} {First-principles theory of the luminescence lineshape for the triplet transition in diamond nv centres},\ }\href {https://doi.org/10.1088/1367-2630/16/7/073026} {\bibfield  {journal} {\bibinfo  {journal} {New Journal of Physics}\ }\textbf {\bibinfo {volume} {16}},\ \bibinfo {pages} {073026} (\bibinfo {year} {2014})}\BibitemShut {NoStop}%
\bibitem [{\citenamefont {Jelezko}\ \emph {et~al.}(2004{\natexlab{a}})\citenamefont {Jelezko}, \citenamefont {Gaebel}, \citenamefont {Popa}, \citenamefont {Gruber},\ and\ \citenamefont {Wrachtrup}}]{NV-SPE3}%
  \BibitemOpen
  \bibfield  {author} {\bibinfo {author} {\bibfnamefont {F.}~\bibnamefont {Jelezko}}, \bibinfo {author} {\bibfnamefont {T.}~\bibnamefont {Gaebel}}, \bibinfo {author} {\bibfnamefont {I.}~\bibnamefont {Popa}}, \bibinfo {author} {\bibfnamefont {A.}~\bibnamefont {Gruber}},\ and\ \bibinfo {author} {\bibfnamefont {J.}~\bibnamefont {Wrachtrup}},\ }\bibfield  {title} {\bibinfo {title} {Observation of coherent oscillations in a single electron spin},\ }\href {https://doi.org/10.1103/PhysRevLett.92.076401} {\bibfield  {journal} {\bibinfo  {journal} {Phys. Rev. Lett.}\ }\textbf {\bibinfo {volume} {92}},\ \bibinfo {pages} {076401} (\bibinfo {year} {2004}{\natexlab{a}})}\BibitemShut {NoStop}%
\bibitem [{\citenamefont {Jelezko}\ \emph {et~al.}(2004{\natexlab{b}})\citenamefont {Jelezko}, \citenamefont {Gaebel}, \citenamefont {Popa}, \citenamefont {Domhan}, \citenamefont {Gruber},\ and\ \citenamefont {Wrachtrup}}]{NV-SPE4}%
  \BibitemOpen
  \bibfield  {author} {\bibinfo {author} {\bibfnamefont {F.}~\bibnamefont {Jelezko}}, \bibinfo {author} {\bibfnamefont {T.}~\bibnamefont {Gaebel}}, \bibinfo {author} {\bibfnamefont {I.}~\bibnamefont {Popa}}, \bibinfo {author} {\bibfnamefont {M.}~\bibnamefont {Domhan}}, \bibinfo {author} {\bibfnamefont {A.}~\bibnamefont {Gruber}},\ and\ \bibinfo {author} {\bibfnamefont {J.}~\bibnamefont {Wrachtrup}},\ }\bibfield  {title} {\bibinfo {title} {Observation of coherent oscillation of a single nuclear spin and realization of a two-qubit conditional quantum gate},\ }\href {https://doi.org/10.1103/PhysRevLett.93.130501} {\bibfield  {journal} {\bibinfo  {journal} {Phys. Rev. Lett.}\ }\textbf {\bibinfo {volume} {93}},\ \bibinfo {pages} {130501} (\bibinfo {year} {2004}{\natexlab{b}})}\BibitemShut {NoStop}%
\bibitem [{\citenamefont {Balasubramanian}\ \emph {et~al.}(2009)\citenamefont {Balasubramanian}, \citenamefont {Neumann}, \citenamefont {Twitchen}, \citenamefont {Markham}, \citenamefont {Kolesov}, \citenamefont {Mizuochi}, \citenamefont {Isoya}, \citenamefont {Achard}, \citenamefont {Beck}, \citenamefont {Tissler}, \citenamefont {Jacques}, \citenamefont {Hemmer},\ and\ \citenamefont {Jelezko}}]{NV-SPE5}%
  \BibitemOpen
  \bibfield  {author} {\bibinfo {author} {\bibfnamefont {G.}~\bibnamefont {Balasubramanian}}, \bibinfo {author} {\bibfnamefont {P.}~\bibnamefont {Neumann}}, \bibinfo {author} {\bibfnamefont {D.}~\bibnamefont {Twitchen}}, \bibinfo {author} {\bibfnamefont {M.}~\bibnamefont {Markham}}, \bibinfo {author} {\bibfnamefont {R.}~\bibnamefont {Kolesov}}, \bibinfo {author} {\bibfnamefont {N.}~\bibnamefont {Mizuochi}}, \bibinfo {author} {\bibfnamefont {J.}~\bibnamefont {Isoya}}, \bibinfo {author} {\bibfnamefont {J.}~\bibnamefont {Achard}}, \bibinfo {author} {\bibfnamefont {J.}~\bibnamefont {Beck}}, \bibinfo {author} {\bibfnamefont {J.}~\bibnamefont {Tissler}}, \bibinfo {author} {\bibfnamefont {V.}~\bibnamefont {Jacques}}, \bibinfo {author} {\bibfnamefont {P.}~\bibnamefont {Hemmer}},\ and\ \bibinfo {author} {\bibfnamefont {F.}~\bibnamefont {Jelezko}},\ }\bibfield  {title} {\bibinfo {title} {Ultralong spin coherence time in isotopically engineered diamond},\ }\href {https://doi.org/10.1038/nmat2420} {\bibfield  {journal}
  {\bibinfo  {journal} {Nature materials}\ }\textbf {\bibinfo {volume} {8}},\ \bibinfo {pages} {383} (\bibinfo {year} {2009})}\BibitemShut {NoStop}%
\bibitem [{\citenamefont {Schell}\ \emph {et~al.}(2017)\citenamefont {Schell}, \citenamefont {Takashima}, \citenamefont {Tran}, \citenamefont {Aharonovich},\ and\ \citenamefont {Takeuchi}}]{TMDCSPE1}%
  \BibitemOpen
  \bibfield  {author} {\bibinfo {author} {\bibfnamefont {A.}~\bibnamefont {Schell}}, \bibinfo {author} {\bibfnamefont {H.}~\bibnamefont {Takashima}}, \bibinfo {author} {\bibfnamefont {T.}~\bibnamefont {Tran}}, \bibinfo {author} {\bibfnamefont {I.}~\bibnamefont {Aharonovich}},\ and\ \bibinfo {author} {\bibfnamefont {S.}~\bibnamefont {Takeuchi}},\ }\bibfield  {title} {\bibinfo {title} {Coupling quantum emitters in 2d materials with tapered fibers},\ }\href {https://doi.org/10.1021/acsphotonics.7b00025} {\bibfield  {journal} {\bibinfo  {journal} {ACS Photonics}\ }\textbf {\bibinfo {volume} {4}} (\bibinfo {year} {2017})}\BibitemShut {NoStop}%
\bibitem [{\citenamefont {Tonndorf}\ \emph {et~al.}(2015)\citenamefont {Tonndorf}, \citenamefont {Schmidt}, \citenamefont {Schneider}, \citenamefont {Kern}, \citenamefont {Buscema}, \citenamefont {Steele}, \citenamefont {Castellanos-Gomez}, \citenamefont {van~der Zant}, \citenamefont {de~Vasconcellos},\ and\ \citenamefont {Bratschitsch}}]{TMDCSPE3}%
  \BibitemOpen
  \bibfield  {author} {\bibinfo {author} {\bibfnamefont {P.}~\bibnamefont {Tonndorf}}, \bibinfo {author} {\bibfnamefont {R.}~\bibnamefont {Schmidt}}, \bibinfo {author} {\bibfnamefont {R.}~\bibnamefont {Schneider}}, \bibinfo {author} {\bibfnamefont {J.}~\bibnamefont {Kern}}, \bibinfo {author} {\bibfnamefont {M.}~\bibnamefont {Buscema}}, \bibinfo {author} {\bibfnamefont {G.~A.}\ \bibnamefont {Steele}}, \bibinfo {author} {\bibfnamefont {A.}~\bibnamefont {Castellanos-Gomez}}, \bibinfo {author} {\bibfnamefont {H.~S.~J.}\ \bibnamefont {van~der Zant}}, \bibinfo {author} {\bibfnamefont {S.~M.}\ \bibnamefont {de~Vasconcellos}},\ and\ \bibinfo {author} {\bibfnamefont {R.}~\bibnamefont {Bratschitsch}},\ }\bibfield  {title} {\bibinfo {title} {Single-photon emission from localized excitons in an atomically thin semiconductor},\ }\href {https://doi.org/10.1364/OPTICA.2.000347} {\bibfield  {journal} {\bibinfo  {journal} {Optica}\ }\textbf {\bibinfo {volume} {2}},\ \bibinfo {pages} {347} (\bibinfo {year} {2015})}\BibitemShut
  {NoStop}%
\bibitem [{\citenamefont {Chakraborty}\ \emph {et~al.}(2015)\citenamefont {Chakraborty}, \citenamefont {Kinnischtzke}, \citenamefont {Goodfellow}, \citenamefont {Beams},\ and\ \citenamefont {Vamivakas}}]{TMDCSPE4}%
  \BibitemOpen
  \bibfield  {author} {\bibinfo {author} {\bibfnamefont {C.}~\bibnamefont {Chakraborty}}, \bibinfo {author} {\bibfnamefont {L.}~\bibnamefont {Kinnischtzke}}, \bibinfo {author} {\bibfnamefont {K.}~\bibnamefont {Goodfellow}}, \bibinfo {author} {\bibfnamefont {R.}~\bibnamefont {Beams}},\ and\ \bibinfo {author} {\bibfnamefont {N.}~\bibnamefont {Vamivakas}},\ }\bibfield  {title} {\bibinfo {title} {Voltage-controlled quantum light from an atomically thin semiconductor},\ }\href {https://doi.org/10.1038/nnano.2015.79} {\bibfield  {journal} {\bibinfo  {journal} {Nature nanotechnology}\ }\textbf {\bibinfo {volume} {10}} (\bibinfo {year} {2015})}\BibitemShut {NoStop}%
\bibitem [{\citenamefont {Palacios-Berraquero}\ \emph {et~al.}(2016)\citenamefont {Palacios-Berraquero}, \citenamefont {Kara}, \citenamefont {Montblanch}, \citenamefont {Barbone}, \citenamefont {Latawiec}, \citenamefont {Yoon}, \citenamefont {Ott}, \citenamefont {Loncar}, \citenamefont {Ferrari},\ and\ \citenamefont {Atatüre}}]{TMDCSPE5}%
  \BibitemOpen
  \bibfield  {author} {\bibinfo {author} {\bibfnamefont {C.}~\bibnamefont {Palacios-Berraquero}}, \bibinfo {author} {\bibfnamefont {D.}~\bibnamefont {Kara}}, \bibinfo {author} {\bibfnamefont {A.}~\bibnamefont {Montblanch}}, \bibinfo {author} {\bibfnamefont {M.}~\bibnamefont {Barbone}}, \bibinfo {author} {\bibfnamefont {P.}~\bibnamefont {Latawiec}}, \bibinfo {author} {\bibfnamefont {D.}~\bibnamefont {Yoon}}, \bibinfo {author} {\bibfnamefont {A.}~\bibnamefont {Ott}}, \bibinfo {author} {\bibfnamefont {M.}~\bibnamefont {Loncar}}, \bibinfo {author} {\bibfnamefont {A.}~\bibnamefont {Ferrari}},\ and\ \bibinfo {author} {\bibfnamefont {M.}~\bibnamefont {Atatüre}},\ }\bibfield  {title} {\bibinfo {title} {Large-scale quantum-emitter arrays in atomically thin semiconductors},\ }\href {https://doi.org/10.1038/ncomms15093} {\bibfield  {journal} {\bibinfo  {journal} {Nature Communications}\ }\textbf {\bibinfo {volume} {8}} (\bibinfo {year} {2016})}\BibitemShut {NoStop}%
\bibitem [{\citenamefont {Branny}\ \emph {et~al.}(2017)\citenamefont {Branny}, \citenamefont {Kumar}, \citenamefont {Proux},\ and\ \citenamefont {Gerardot}}]{TMDCSPE6}%
  \BibitemOpen
  \bibfield  {author} {\bibinfo {author} {\bibfnamefont {A.}~\bibnamefont {Branny}}, \bibinfo {author} {\bibfnamefont {S.}~\bibnamefont {Kumar}}, \bibinfo {author} {\bibfnamefont {R.}~\bibnamefont {Proux}},\ and\ \bibinfo {author} {\bibfnamefont {B.}~\bibnamefont {Gerardot}},\ }\bibfield  {title} {\bibinfo {title} {Deterministic strain-induced arrays of quantum emitters in a two-dimensional semiconductor},\ }\href {https://doi.org/10.1038/ncomms15053} {\bibfield  {journal} {\bibinfo  {journal} {Nature Communications}\ }\textbf {\bibinfo {volume} {8}},\ \bibinfo {pages} {15053} (\bibinfo {year} {2017})}\BibitemShut {NoStop}%
\bibitem [{\citenamefont {Wu}\ \emph {et~al.}(2019{\natexlab{b}})\citenamefont {Wu}, \citenamefont {Dass}, \citenamefont {Hendrickson}, \citenamefont {Montaño}, \citenamefont {Fischer}, \citenamefont {Zhang}, \citenamefont {Choudhury}, \citenamefont {Redwing}, \citenamefont {Wang},\ and\ \citenamefont {Pettes}}]{TMDCSPE7}%
  \BibitemOpen
  \bibfield  {author} {\bibinfo {author} {\bibfnamefont {W.}~\bibnamefont {Wu}}, \bibinfo {author} {\bibfnamefont {C.~K.}\ \bibnamefont {Dass}}, \bibinfo {author} {\bibfnamefont {J.~R.}\ \bibnamefont {Hendrickson}}, \bibinfo {author} {\bibfnamefont {R.~D.}\ \bibnamefont {Montaño}}, \bibinfo {author} {\bibfnamefont {R.~E.}\ \bibnamefont {Fischer}}, \bibinfo {author} {\bibfnamefont {X.}~\bibnamefont {Zhang}}, \bibinfo {author} {\bibfnamefont {T.~H.}\ \bibnamefont {Choudhury}}, \bibinfo {author} {\bibfnamefont {J.~M.}\ \bibnamefont {Redwing}}, \bibinfo {author} {\bibfnamefont {Y.}~\bibnamefont {Wang}},\ and\ \bibinfo {author} {\bibfnamefont {M.~T.}\ \bibnamefont {Pettes}},\ }\bibfield  {title} {\bibinfo {title} {Locally defined quantum emission from epitaxial few-layer tungsten diselenide},\ }\href {https://doi.org/10.1063/1.5091779} {\bibfield  {journal} {\bibinfo  {journal} {Applied Physics Letters}\ }\textbf {\bibinfo {volume} {114}},\ \bibinfo {pages} {213102} (\bibinfo {year} {2019}{\natexlab{b}})},\
  \Eprint {https://arxiv.org/abs/https://pubs.aip.org/aip/apl/article-pdf/doi/10.1063/1.5091779/13419608/213102\_1\_online.pdf} {https://pubs.aip.org/aip/apl/article-pdf/doi/10.1063/1.5091779/13419608/213102\_1\_online.pdf} \BibitemShut {NoStop}%
\bibitem [{\citenamefont {He}\ \emph {et~al.}(2014)\citenamefont {He}, \citenamefont {Clark}, \citenamefont {Schaibley}, \citenamefont {He}, \citenamefont {Chen}, \citenamefont {Wei}, \citenamefont {Ding}, \citenamefont {Zhang}, \citenamefont {Yao}, \citenamefont {Xu}, \citenamefont {Lu},\ and\ \citenamefont {Pan}}]{TMDCSPE8}%
  \BibitemOpen
  \bibfield  {author} {\bibinfo {author} {\bibfnamefont {Y.-M.}\ \bibnamefont {He}}, \bibinfo {author} {\bibfnamefont {G.}~\bibnamefont {Clark}}, \bibinfo {author} {\bibfnamefont {J.}~\bibnamefont {Schaibley}}, \bibinfo {author} {\bibfnamefont {Y.}~\bibnamefont {He}}, \bibinfo {author} {\bibfnamefont {M.-c.}\ \bibnamefont {Chen}}, \bibinfo {author} {\bibfnamefont {Y.-J.}\ \bibnamefont {Wei}}, \bibinfo {author} {\bibfnamefont {X.}~\bibnamefont {Ding}}, \bibinfo {author} {\bibfnamefont {Q.}~\bibnamefont {Zhang}}, \bibinfo {author} {\bibfnamefont {W.}~\bibnamefont {Yao}}, \bibinfo {author} {\bibfnamefont {X.}~\bibnamefont {Xu}}, \bibinfo {author} {\bibfnamefont {C.-Y.}\ \bibnamefont {Lu}},\ and\ \bibinfo {author} {\bibfnamefont {J.-W.}\ \bibnamefont {Pan}},\ }\bibfield  {title} {\bibinfo {title} {Single quantum emitters in monolayer semiconductors},\ }\href {https://doi.org/10.1038/nnano.2015.75} {\bibfield  {journal} {\bibinfo  {journal} {Nature nanotechnology}\ }\textbf {\bibinfo {volume} {10}} (\bibinfo
  {year} {2014})}\BibitemShut {NoStop}%
\bibitem [{\citenamefont {Ouma}\ \emph {et~al.}(2017)\citenamefont {Ouma}, \citenamefont {Singh}, \citenamefont {Obodo}, \citenamefont {Amolo},\ and\ \citenamefont {Romero}}]{LnTMDCSPE1}%
  \BibitemOpen
  \bibfield  {author} {\bibinfo {author} {\bibfnamefont {C.~N.~M.}\ \bibnamefont {Ouma}}, \bibinfo {author} {\bibfnamefont {S.}~\bibnamefont {Singh}}, \bibinfo {author} {\bibfnamefont {K.~O.}\ \bibnamefont {Obodo}}, \bibinfo {author} {\bibfnamefont {G.~O.}\ \bibnamefont {Amolo}},\ and\ \bibinfo {author} {\bibfnamefont {A.~H.}\ \bibnamefont {Romero}},\ }\bibfield  {title} {\bibinfo {title} {Controlling the magnetic and optical responses of a mos2 monolayer by lanthanide substitutional doping: a first-principles study},\ }\href {https://doi.org/10.1039/C7CP03160B} {\bibfield  {journal} {\bibinfo  {journal} {Phys. Chem. Chem. Phys.}\ }\textbf {\bibinfo {volume} {19}},\ \bibinfo {pages} {25555} (\bibinfo {year} {2017})}\BibitemShut {NoStop}%
\bibitem [{\citenamefont {Li}\ \emph {et~al.}(2021)\citenamefont {Li}, \citenamefont {Tian}, \citenamefont {Yao}, \citenamefont {He}, \citenamefont {Chen}, \citenamefont {Zhang},\ and\ \citenamefont {Zhai}}]{LnTMDCSPE2}%
  \BibitemOpen
  \bibfield  {author} {\bibinfo {author} {\bibfnamefont {S.}~\bibnamefont {Li}}, \bibinfo {author} {\bibfnamefont {S.}~\bibnamefont {Tian}}, \bibinfo {author} {\bibfnamefont {Y.}~\bibnamefont {Yao}}, \bibinfo {author} {\bibfnamefont {M.}~\bibnamefont {He}}, \bibinfo {author} {\bibfnamefont {L.}~\bibnamefont {Chen}}, \bibinfo {author} {\bibfnamefont {Y.}~\bibnamefont {Zhang}},\ and\ \bibinfo {author} {\bibfnamefont {J.}~\bibnamefont {Zhai}},\ }\bibfield  {title} {\bibinfo {title} {Enhanced electrical performance of monolayer mos2 with rare earth element sm doping},\ }\href {https://doi.org/10.3390/nano11030769} {\bibfield  {journal} {\bibinfo  {journal} {Nanomaterials}\ }\textbf {\bibinfo {volume} {11}},\ \bibinfo {pages} {769} (\bibinfo {year} {2021})}\BibitemShut {NoStop}%
\bibitem [{\citenamefont {López-Morales}\ \emph {et~al.}(2021)\citenamefont {López-Morales}, \citenamefont {Hampel}, \citenamefont {López}, \citenamefont {Menon}, \citenamefont {Flick},\ and\ \citenamefont {Meriles}}]{LnTMDCSPE3_WS2}%
  \BibitemOpen
  \bibfield  {author} {\bibinfo {author} {\bibfnamefont {G.}~\bibnamefont {López-Morales}}, \bibinfo {author} {\bibfnamefont {A.}~\bibnamefont {Hampel}}, \bibinfo {author} {\bibfnamefont {G.}~\bibnamefont {López}}, \bibinfo {author} {\bibfnamefont {V.}~\bibnamefont {Menon}}, \bibinfo {author} {\bibfnamefont {J.}~\bibnamefont {Flick}},\ and\ \bibinfo {author} {\bibfnamefont {C.}~\bibnamefont {Meriles}},\ }\bibfield  {title} {\bibinfo {title} {Ab-initio investigation of er3+ defects in tungsten disulfide},\ }\href {https://doi.org/10.1016/j.commatsci.2021.111041} {\bibfield  {journal} {\bibinfo  {journal} {Computational Materials Science}\ }\textbf {\bibinfo {volume} {210}},\ \bibinfo {pages} {111041} (\bibinfo {year} {2021})}\BibitemShut {NoStop}%
\bibitem [{\citenamefont {Maleki-Ghaleh}\ \emph {et~al.}(2024)\citenamefont {Maleki-Ghaleh}, \citenamefont {Moradpur-Tari}, \citenamefont {Shakiba}, \citenamefont {Paczesny}, \citenamefont {Hurley}, \citenamefont {Siadati}, \citenamefont {Ansari},\ and\ \citenamefont {Gity}}]{LnTMDCSPE4}%
  \BibitemOpen
  \bibfield  {author} {\bibinfo {author} {\bibfnamefont {H.}~\bibnamefont {Maleki-Ghaleh}}, \bibinfo {author} {\bibfnamefont {E.}~\bibnamefont {Moradpur-Tari}}, \bibinfo {author} {\bibfnamefont {M.}~\bibnamefont {Shakiba}}, \bibinfo {author} {\bibfnamefont {J.}~\bibnamefont {Paczesny}}, \bibinfo {author} {\bibfnamefont {P.~K.}\ \bibnamefont {Hurley}}, \bibinfo {author} {\bibfnamefont {M.~H.}\ \bibnamefont {Siadati}}, \bibinfo {author} {\bibfnamefont {L.}~\bibnamefont {Ansari}},\ and\ \bibinfo {author} {\bibfnamefont {F.}~\bibnamefont {Gity}},\ }\bibfield  {title} {\bibinfo {title} {Electronic structure of rare-earth erbium-doped platinum diselenide: A density functional theory study},\ }\href {https://doi.org/https://doi.org/10.1016/j.jpcs.2024.112004} {\bibfield  {journal} {\bibinfo  {journal} {Journal of Physics and Chemistry of Solids}\ }\textbf {\bibinfo {volume} {190}},\ \bibinfo {pages} {112004} (\bibinfo {year} {2024})}\BibitemShut {NoStop}%
\bibitem [{\citenamefont {Dejneka}\ and\ \citenamefont {Samson}(1999)}]{Lnusage1}%
  \BibitemOpen
  \bibfield  {author} {\bibinfo {author} {\bibfnamefont {M.~J.}\ \bibnamefont {Dejneka}}\ and\ \bibinfo {author} {\bibfnamefont {B.}~\bibnamefont {Samson}},\ }\bibfield  {title} {\bibinfo {title} {Rare-earth-doped fibers for telecommunications applications},\ }\href {https://api.semanticscholar.org/CorpusID:136117491} {\bibfield  {journal} {\bibinfo  {journal} {MRS Bulletin}\ }\textbf {\bibinfo {volume} {24}},\ \bibinfo {pages} {39} (\bibinfo {year} {1999})}\BibitemShut {NoStop}%
\bibitem [{\citenamefont {Tanabe}(2002)}]{Lnusage2}%
  \BibitemOpen
  \bibfield  {author} {\bibinfo {author} {\bibfnamefont {S.}~\bibnamefont {Tanabe}},\ }\bibfield  {title} {\bibinfo {title} {Rare-earth-doped glasses for fiber amplifiers in broadband telecommunication},\ }\href {https://doi.org/10.1016/S1631-0748(02)01449-2} {\bibfield  {journal} {\bibinfo  {journal} {Comptes Rendus Chimie - C R CHIM}\ }\textbf {\bibinfo {volume} {5}},\ \bibinfo {pages} {815} (\bibinfo {year} {2002})}\BibitemShut {NoStop}%
\bibitem [{\citenamefont {Bünzli}(2017)}]{Lnusage3}%
  \BibitemOpen
  \bibfield  {author} {\bibinfo {author} {\bibfnamefont {J.-C.~G.}\ \bibnamefont {Bünzli}},\ }\bibfield  {title} {\bibinfo {title} {Rising stars in science and technology: Luminescent lanthanide materials},\ }\href {https://doi.org/https://doi.org/10.1002/ejic.201701201} {\bibfield  {journal} {\bibinfo  {journal} {European Journal of Inorganic Chemistry}\ }\textbf {\bibinfo {volume} {2017}},\ \bibinfo {pages} {5058} (\bibinfo {year} {2017})},\ \Eprint {https://arxiv.org/abs/https://chemistry-europe.onlinelibrary.wiley.com/doi/pdf/10.1002/ejic.201701201} {https://chemistry-europe.onlinelibrary.wiley.com/doi/pdf/10.1002/ejic.201701201} \BibitemShut {NoStop}%
\bibitem [{\citenamefont {Stevenson}\ \emph {et~al.}(2022)\citenamefont {Stevenson}, \citenamefont {Phenicie}, \citenamefont {Gray}, \citenamefont {Horvath}, \citenamefont {Welinski}, \citenamefont {Ferrenti}, \citenamefont {Ferrier}, \citenamefont {Goldner}, \citenamefont {Das}, \citenamefont {Ramesh}, \citenamefont {Cava}, \citenamefont {de~Leon},\ and\ \citenamefont {Thompson}}]{Lnusage4}%
  \BibitemOpen
  \bibfield  {author} {\bibinfo {author} {\bibfnamefont {P.}~\bibnamefont {Stevenson}}, \bibinfo {author} {\bibfnamefont {C.~M.}\ \bibnamefont {Phenicie}}, \bibinfo {author} {\bibfnamefont {I.}~\bibnamefont {Gray}}, \bibinfo {author} {\bibfnamefont {S.~P.}\ \bibnamefont {Horvath}}, \bibinfo {author} {\bibfnamefont {S.}~\bibnamefont {Welinski}}, \bibinfo {author} {\bibfnamefont {A.~M.}\ \bibnamefont {Ferrenti}}, \bibinfo {author} {\bibfnamefont {A.}~\bibnamefont {Ferrier}}, \bibinfo {author} {\bibfnamefont {P.}~\bibnamefont {Goldner}}, \bibinfo {author} {\bibfnamefont {S.}~\bibnamefont {Das}}, \bibinfo {author} {\bibfnamefont {R.}~\bibnamefont {Ramesh}}, \bibinfo {author} {\bibfnamefont {R.~J.}\ \bibnamefont {Cava}}, \bibinfo {author} {\bibfnamefont {N.~P.}\ \bibnamefont {de~Leon}},\ and\ \bibinfo {author} {\bibfnamefont {J.~D.}\ \bibnamefont {Thompson}},\ }\bibfield  {title} {\bibinfo {title} {Erbium-implanted materials for quantum communication applications},\ }\href
  {https://doi.org/10.1103/PhysRevB.105.224106} {\bibfield  {journal} {\bibinfo  {journal} {Phys. Rev. B}\ }\textbf {\bibinfo {volume} {105}},\ \bibinfo {pages} {224106} (\bibinfo {year} {2022})}\BibitemShut {NoStop}%
\bibitem [{\citenamefont {Shannon}(1976)}]{Ionicradii}%
  \BibitemOpen
  \bibfield  {author} {\bibinfo {author} {\bibfnamefont {R.}~\bibnamefont {Shannon}},\ }\bibfield  {title} {\bibinfo {title} {Revised effective ionic radii and systematic studies of interatomic distances in halides and chalcogenides},\ }\href@noop {} {\bibfield  {journal} {\bibinfo  {journal} {Acta Cryst}\ }\textbf {\bibinfo {volume} {32}},\ \bibinfo {pages} {751} (\bibinfo {year} {1976})}\BibitemShut {NoStop}%
\bibitem [{\citenamefont {Noh}\ \emph {et~al.}(2014)\citenamefont {Noh}, \citenamefont {Kim},\ and\ \citenamefont {Kim}}]{VSMoS21_Eform_VMoPCHGD}%
  \BibitemOpen
  \bibfield  {author} {\bibinfo {author} {\bibfnamefont {J.-Y.}\ \bibnamefont {Noh}}, \bibinfo {author} {\bibfnamefont {H.}~\bibnamefont {Kim}},\ and\ \bibinfo {author} {\bibfnamefont {Y.-S.}\ \bibnamefont {Kim}},\ }\bibfield  {title} {\bibinfo {title} {Stability and electronic structures of native defects in single-{layerMoS2}},\ }\href {https://doi.org/10.1103/physrevb.89.205417} {\bibfield  {journal} {\bibinfo  {journal} {Phys. Rev. B Condens. Matter Mater. Phys.}\ }\textbf {\bibinfo {volume} {89}},\ \bibinfo {pages} {205417} (\bibinfo {year} {2014})}\BibitemShut {NoStop}%
\bibitem [{\citenamefont {Komsa}\ \emph {et~al.}(2012)\citenamefont {Komsa}, \citenamefont {Kotakoski}, \citenamefont {Kurasch}, \citenamefont {Lehtinen}, \citenamefont {Kaiser},\ and\ \citenamefont {Krasheninnikov}}]{VSMoS22}%
  \BibitemOpen
  \bibfield  {author} {\bibinfo {author} {\bibfnamefont {H.-P.}\ \bibnamefont {Komsa}}, \bibinfo {author} {\bibfnamefont {J.}~\bibnamefont {Kotakoski}}, \bibinfo {author} {\bibfnamefont {S.}~\bibnamefont {Kurasch}}, \bibinfo {author} {\bibfnamefont {O.}~\bibnamefont {Lehtinen}}, \bibinfo {author} {\bibfnamefont {U.}~\bibnamefont {Kaiser}},\ and\ \bibinfo {author} {\bibfnamefont {A.~V.}\ \bibnamefont {Krasheninnikov}},\ }\bibfield  {title} {\bibinfo {title} {Two-dimensional transition metal dichalcogenides under electron irradiation: Defect production and doping},\ }\href {https://doi.org/10.1103/PhysRevLett.109.035503} {\bibfield  {journal} {\bibinfo  {journal} {Phys. Rev. Lett.}\ }\textbf {\bibinfo {volume} {109}},\ \bibinfo {pages} {035503} (\bibinfo {year} {2012})}\BibitemShut {NoStop}%
\bibitem [{\citenamefont {Liu}\ \emph {et~al.}(2013)\citenamefont {Liu}, \citenamefont {Guo}, \citenamefont {Fang},\ and\ \citenamefont {Robertson}}]{VSMoS23_Eform}%
  \BibitemOpen
  \bibfield  {author} {\bibinfo {author} {\bibfnamefont {D.}~\bibnamefont {Liu}}, \bibinfo {author} {\bibfnamefont {Y.}~\bibnamefont {Guo}}, \bibinfo {author} {\bibfnamefont {L.}~\bibnamefont {Fang}},\ and\ \bibinfo {author} {\bibfnamefont {J.}~\bibnamefont {Robertson}},\ }\bibfield  {title} {\bibinfo {title} {Sulfur vacancies in monolayer mos2 and its electrical contacts},\ }\href {https://doi.org/10.1063/1.4824893} {\bibfield  {journal} {\bibinfo  {journal} {Applied Physics Letters}\ }\textbf {\bibinfo {volume} {103}},\ \bibinfo {pages} {183113} (\bibinfo {year} {2013})},\ \Eprint {https://arxiv.org/abs/https://pubs.aip.org/aip/apl/article-pdf/doi/10.1063/1.4824893/13322476/183113\_1\_online.pdf} {https://pubs.aip.org/aip/apl/article-pdf/doi/10.1063/1.4824893/13322476/183113\_1\_online.pdf} \BibitemShut {NoStop}%
\bibitem [{\citenamefont {Salehi}\ and\ \citenamefont {Saffarzadeh}(2016)}]{VSMoS24}%
  \BibitemOpen
  \bibfield  {author} {\bibinfo {author} {\bibfnamefont {S.}~\bibnamefont {Salehi}}\ and\ \bibinfo {author} {\bibfnamefont {A.}~\bibnamefont {Saffarzadeh}},\ }\bibfield  {title} {\bibinfo {title} {Atomic defect states in monolayers of mos2 and ws2},\ }\href {https://doi.org/https://doi.org/10.1016/j.susc.2016.05.003} {\bibfield  {journal} {\bibinfo  {journal} {Surface Science}\ }\textbf {\bibinfo {volume} {651}},\ \bibinfo {pages} {215} (\bibinfo {year} {2016})}\BibitemShut {NoStop}%
\bibitem [{\citenamefont {Komsa}\ and\ \citenamefont {Krasheninnikov}(2015)}]{VSMoS25_Eform}%
  \BibitemOpen
  \bibfield  {author} {\bibinfo {author} {\bibfnamefont {H.-P.}\ \bibnamefont {Komsa}}\ and\ \bibinfo {author} {\bibfnamefont {A.~V.}\ \bibnamefont {Krasheninnikov}},\ }\bibfield  {title} {\bibinfo {title} {Native defects in bulk and monolayer ${\mathrm{mos}}_{2}$ from first principles},\ }\href {https://doi.org/10.1103/PhysRevB.91.125304} {\bibfield  {journal} {\bibinfo  {journal} {Phys. Rev. B}\ }\textbf {\bibinfo {volume} {91}},\ \bibinfo {pages} {125304} (\bibinfo {year} {2015})}\BibitemShut {NoStop}%
\bibitem [{\citenamefont {Zhou}\ \emph {et~al.}(2013)\citenamefont {Zhou}, \citenamefont {Zou}, \citenamefont {Najmaei}, \citenamefont {Liu}, \citenamefont {Shi}, \citenamefont {Kong}, \citenamefont {Lou}, \citenamefont {Ajayan}, \citenamefont {Yakobson},\ and\ \citenamefont {Idrobo}}]{VSMoS26_Eform}%
  \BibitemOpen
  \bibfield  {author} {\bibinfo {author} {\bibfnamefont {W.}~\bibnamefont {Zhou}}, \bibinfo {author} {\bibfnamefont {X.}~\bibnamefont {Zou}}, \bibinfo {author} {\bibfnamefont {S.}~\bibnamefont {Najmaei}}, \bibinfo {author} {\bibfnamefont {Z.}~\bibnamefont {Liu}}, \bibinfo {author} {\bibfnamefont {Y.}~\bibnamefont {Shi}}, \bibinfo {author} {\bibfnamefont {J.}~\bibnamefont {Kong}}, \bibinfo {author} {\bibfnamefont {J.}~\bibnamefont {Lou}}, \bibinfo {author} {\bibfnamefont {P.~M.}\ \bibnamefont {Ajayan}}, \bibinfo {author} {\bibfnamefont {B.~I.}\ \bibnamefont {Yakobson}},\ and\ \bibinfo {author} {\bibfnamefont {J.-C.}\ \bibnamefont {Idrobo}},\ }\bibfield  {title} {\bibinfo {title} {Intrinsic structural defects in monolayer molybdenum disulfide},\ }\href {https://doi.org/10.1021/nl4007479} {\bibfield  {journal} {\bibinfo  {journal} {Nano Letters}\ }\textbf {\bibinfo {volume} {13}},\ \bibinfo {pages} {2615} (\bibinfo {year} {2013})}\BibitemShut {NoStop}%
\bibitem [{\citenamefont {Qiu}\ \emph {et~al.}(2013)\citenamefont {Qiu}, \citenamefont {Xu}, \citenamefont {Wang}, \citenamefont {Ren}, \citenamefont {Nan}, \citenamefont {Ni}, \citenamefont {Chen}, \citenamefont {Yuan}, \citenamefont {Miao}, \citenamefont {Song}, \citenamefont {Long}, \citenamefont {Shi}, \citenamefont {Sun}, \citenamefont {Wang},\ and\ \citenamefont {Wang}}]{VSMoS27}%
  \BibitemOpen
  \bibfield  {author} {\bibinfo {author} {\bibfnamefont {H.}~\bibnamefont {Qiu}}, \bibinfo {author} {\bibfnamefont {T.}~\bibnamefont {Xu}}, \bibinfo {author} {\bibfnamefont {Z.}~\bibnamefont {Wang}}, \bibinfo {author} {\bibfnamefont {W.}~\bibnamefont {Ren}}, \bibinfo {author} {\bibfnamefont {H.}~\bibnamefont {Nan}}, \bibinfo {author} {\bibfnamefont {Z.}~\bibnamefont {Ni}}, \bibinfo {author} {\bibfnamefont {Q.}~\bibnamefont {Chen}}, \bibinfo {author} {\bibfnamefont {S.}~\bibnamefont {Yuan}}, \bibinfo {author} {\bibfnamefont {F.}~\bibnamefont {Miao}}, \bibinfo {author} {\bibfnamefont {F.}~\bibnamefont {Song}}, \bibinfo {author} {\bibfnamefont {G.}~\bibnamefont {Long}}, \bibinfo {author} {\bibfnamefont {Y.}~\bibnamefont {Shi}}, \bibinfo {author} {\bibfnamefont {L.}~\bibnamefont {Sun}}, \bibinfo {author} {\bibfnamefont {J.}~\bibnamefont {Wang}},\ and\ \bibinfo {author} {\bibfnamefont {X.}~\bibnamefont {Wang}},\ }\bibfield  {title} {\bibinfo {title} {Hopping transport through defect-induced localized states in
  molybdenum disulphide},\ }\href {https://doi.org/10.1038/ncomms3642} {\bibfield  {journal} {\bibinfo  {journal} {Nature communications}\ }\textbf {\bibinfo {volume} {4}},\ \bibinfo {pages} {2642} (\bibinfo {year} {2013})}\BibitemShut {NoStop}%
\bibitem [{\citenamefont {Kresse}\ and\ \citenamefont {Furthm\"uller}(1996{\natexlab{a}})}]{Method_VASP}%
  \BibitemOpen
  \bibfield  {author} {\bibinfo {author} {\bibfnamefont {G.}~\bibnamefont {Kresse}}\ and\ \bibinfo {author} {\bibfnamefont {J.}~\bibnamefont {Furthm\"uller}},\ }\bibfield  {title} {\bibinfo {title} {Efficiency of ab-initio total energy calculations for metals and semiconductors using a plane-wave basis set},\ }\href {https://doi.org/10.1016/0927-0256(96)00008-0} {\bibfield  {journal} {\bibinfo  {journal} {Comput. Mater. Sci.}\ }\textbf {\bibinfo {volume} {6}},\ \bibinfo {pages} {15 } (\bibinfo {year} {1996}{\natexlab{a}})}\BibitemShut {NoStop}%
\bibitem [{\citenamefont {Perdew}\ \emph {et~al.}(1996)\citenamefont {Perdew}, \citenamefont {Burke},\ and\ \citenamefont {Ernzerhof}}]{Method_GGA}%
  \BibitemOpen
  \bibfield  {author} {\bibinfo {author} {\bibfnamefont {J.~P.}\ \bibnamefont {Perdew}}, \bibinfo {author} {\bibfnamefont {K.}~\bibnamefont {Burke}},\ and\ \bibinfo {author} {\bibfnamefont {M.}~\bibnamefont {Ernzerhof}},\ }\bibfield  {title} {\bibinfo {title} {Generalized gradient approximation made simple},\ }\href {https://doi.org/10.1103/PhysRevLett.77.3865} {\bibfield  {journal} {\bibinfo  {journal} {Phys. Rev. Lett.}\ }\textbf {\bibinfo {volume} {77}},\ \bibinfo {pages} {3865} (\bibinfo {year} {1996})}\BibitemShut {NoStop}%
\bibitem [{\citenamefont {Fuchs}\ and\ \citenamefont {Scheffler}(1999)}]{Method_PAW}%
  \BibitemOpen
  \bibfield  {author} {\bibinfo {author} {\bibfnamefont {M.}~\bibnamefont {Fuchs}}\ and\ \bibinfo {author} {\bibfnamefont {M.}~\bibnamefont {Scheffler}},\ }\bibfield  {title} {\bibinfo {title} {Ab initio pseudopotentials for electronic structure calculations of poly-atomic systems using density-functional theory},\ }\href {https://doi.org/10.1016/S0010-4655(98)00201-X} {\bibfield  {journal} {\bibinfo  {journal} {Computer Physics Communications}\ }\textbf {\bibinfo {volume} {119}},\ \bibinfo {pages} {67} (\bibinfo {year} {1999})}\BibitemShut {NoStop}%
\bibitem [{\citenamefont {Perdew}\ \emph {et~al.}(1998)\citenamefont {Perdew}, \citenamefont {Burke},\ and\ \citenamefont {Ernzerhof}}]{Method_PBE}%
  \BibitemOpen
  \bibfield  {author} {\bibinfo {author} {\bibfnamefont {J.~P.}\ \bibnamefont {Perdew}}, \bibinfo {author} {\bibfnamefont {K.}~\bibnamefont {Burke}},\ and\ \bibinfo {author} {\bibfnamefont {M.}~\bibnamefont {Ernzerhof}},\ }\bibfield  {title} {\bibinfo {title} {Perdew, burke, and ernzerhof reply:},\ }\href {https://doi.org/10.1103/PhysRevLett.80.891} {\bibfield  {journal} {\bibinfo  {journal} {Phys. Rev. Lett.}\ }\textbf {\bibinfo {volume} {80}},\ \bibinfo {pages} {891} (\bibinfo {year} {1998})}\BibitemShut {NoStop}%
\bibitem [{\citenamefont {Kresse}\ and\ \citenamefont {Furthm\"uller}(1996{\natexlab{b}})}]{Method_planewavebasisset}%
  \BibitemOpen
  \bibfield  {author} {\bibinfo {author} {\bibfnamefont {G.}~\bibnamefont {Kresse}}\ and\ \bibinfo {author} {\bibfnamefont {J.}~\bibnamefont {Furthm\"uller}},\ }\bibfield  {title} {\bibinfo {title} {Efficient iterative schemes for ab initio total-energy calculations using a plane-wave basis set},\ }\href {https://doi.org/10.1103/PhysRevB.54.11169} {\bibfield  {journal} {\bibinfo  {journal} {Phys. Rev. B}\ }\textbf {\bibinfo {volume} {54}},\ \bibinfo {pages} {11169} (\bibinfo {year} {1996}{\natexlab{b}})}\BibitemShut {NoStop}%
\bibitem [{\citenamefont {Chagas~da Silva}\ \emph {et~al.}(2021)\citenamefont {Chagas~da Silva}, \citenamefont {Lorke}, \citenamefont {Aradi}, \citenamefont {Farzalipour~Tabriz}, \citenamefont {Frauenheim}, \citenamefont {Rubio}, \citenamefont {Rocca},\ and\ \citenamefont {De\'ak}}]{Method_SCPC}%
  \BibitemOpen
  \bibfield  {author} {\bibinfo {author} {\bibfnamefont {M.}~\bibnamefont {Chagas~da Silva}}, \bibinfo {author} {\bibfnamefont {M.}~\bibnamefont {Lorke}}, \bibinfo {author} {\bibfnamefont {B.}~\bibnamefont {Aradi}}, \bibinfo {author} {\bibfnamefont {M.}~\bibnamefont {Farzalipour~Tabriz}}, \bibinfo {author} {\bibfnamefont {T.}~\bibnamefont {Frauenheim}}, \bibinfo {author} {\bibfnamefont {A.}~\bibnamefont {Rubio}}, \bibinfo {author} {\bibfnamefont {D.}~\bibnamefont {Rocca}},\ and\ \bibinfo {author} {\bibfnamefont {P.}~\bibnamefont {De\'ak}},\ }\bibfield  {title} {\bibinfo {title} {Self-consistent potential correction for charged periodic systems},\ }\href {https://doi.org/10.1103/PhysRevLett.126.076401} {\bibfield  {journal} {\bibinfo  {journal} {Phys. Rev. Lett.}\ }\textbf {\bibinfo {volume} {126}},\ \bibinfo {pages} {076401} (\bibinfo {year} {2021})}\BibitemShut {NoStop}%
\bibitem [{\citenamefont {Mak}\ \emph {et~al.}(2010)\citenamefont {Mak}, \citenamefont {Lee}, \citenamefont {Hone}, \citenamefont {Shan},\ and\ \citenamefont {Heinz}}]{MoS2Eg1}%
  \BibitemOpen
  \bibfield  {author} {\bibinfo {author} {\bibfnamefont {K.~F.}\ \bibnamefont {Mak}}, \bibinfo {author} {\bibfnamefont {C.}~\bibnamefont {Lee}}, \bibinfo {author} {\bibfnamefont {J.}~\bibnamefont {Hone}}, \bibinfo {author} {\bibfnamefont {J.}~\bibnamefont {Shan}},\ and\ \bibinfo {author} {\bibfnamefont {T.~F.}\ \bibnamefont {Heinz}},\ }\bibfield  {title} {\bibinfo {title} {Atomically thin ${\mathrm{mos}}_{2}$: A new direct-gap semiconductor},\ }\href {https://doi.org/10.1103/PhysRevLett.105.136805} {\bibfield  {journal} {\bibinfo  {journal} {Phys. Rev. Lett.}\ }\textbf {\bibinfo {volume} {105}},\ \bibinfo {pages} {136805} (\bibinfo {year} {2010})}\BibitemShut {NoStop}%
\bibitem [{\citenamefont {Splendiani}\ \emph {et~al.}(2010)\citenamefont {Splendiani}, \citenamefont {Sun}, \citenamefont {Zhang}, \citenamefont {Li}, \citenamefont {Kim}, \citenamefont {Chim}, \citenamefont {Galli},\ and\ \citenamefont {Wang}}]{MoS2Eg2}%
  \BibitemOpen
  \bibfield  {author} {\bibinfo {author} {\bibfnamefont {A.}~\bibnamefont {Splendiani}}, \bibinfo {author} {\bibfnamefont {L.}~\bibnamefont {Sun}}, \bibinfo {author} {\bibfnamefont {Y.}~\bibnamefont {Zhang}}, \bibinfo {author} {\bibfnamefont {T.}~\bibnamefont {Li}}, \bibinfo {author} {\bibfnamefont {J.}~\bibnamefont {Kim}}, \bibinfo {author} {\bibfnamefont {C.-Y.}\ \bibnamefont {Chim}}, \bibinfo {author} {\bibfnamefont {G.}~\bibnamefont {Galli}},\ and\ \bibinfo {author} {\bibfnamefont {F.}~\bibnamefont {Wang}},\ }\bibfield  {title} {\bibinfo {title} {Emerging photoluminescence in monolayer mos2},\ }\href {https://doi.org/10.1021/nl903868w} {\bibfield  {journal} {\bibinfo  {journal} {Nano Letters}\ }\textbf {\bibinfo {volume} {10}},\ \bibinfo {pages} {1271} (\bibinfo {year} {2010})}\BibitemShut {NoStop}%
\bibitem [{\citenamefont {Kadantsev}\ and\ \citenamefont {Hawrylak}(2012)}]{MoS2Eg3}%
  \BibitemOpen
  \bibfield  {author} {\bibinfo {author} {\bibfnamefont {E.~S.}\ \bibnamefont {Kadantsev}}\ and\ \bibinfo {author} {\bibfnamefont {P.}~\bibnamefont {Hawrylak}},\ }\bibfield  {title} {\bibinfo {title} {Electronic structure of a single mos2 monolayer},\ }\href {https://doi.org/https://doi.org/10.1016/j.ssc.2012.02.005} {\bibfield  {journal} {\bibinfo  {journal} {Solid State Communications}\ }\textbf {\bibinfo {volume} {152}},\ \bibinfo {pages} {909} (\bibinfo {year} {2012})}\BibitemShut {NoStop}%
\end{thebibliography}%
\clearpage
\onecolumngrid
\appendix
\section*{Supplemental Figures}
\counterwithin{figure}{section}
\renewcommand\thefigure{App.\arabic{figure}}
\begin{figure*}[ht]
    \centering
    \includegraphics[width=0.4\textwidth]{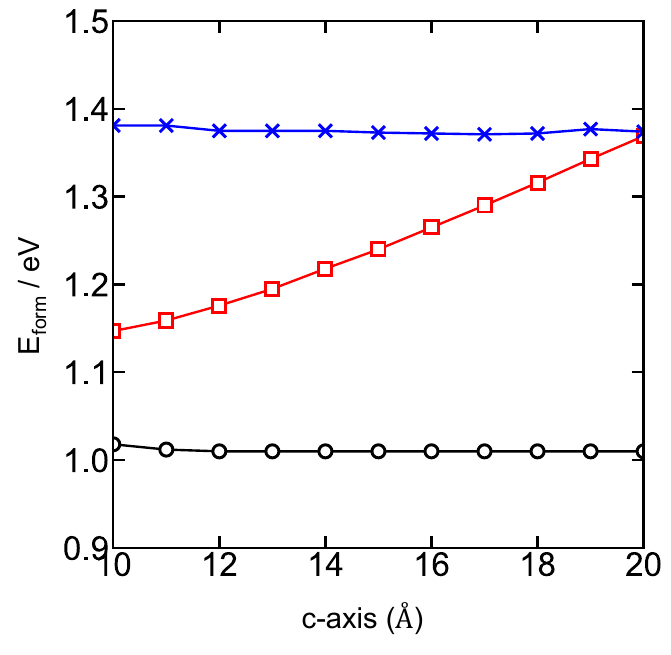}
    \caption{Formation energies of \defect{Ce}{Mo} as a function of c-axis. The black line with circles corresponds to neutral \defect{Ce}{Mo}. The red line with squares corresponds to negatively charged \defect{Ce}{Mo} without consideration of the energy correction term ($E_{corr}$). The blue line with crosses corresponds to negatively charged \defect{Ce}{Mo} with consideration of the energy correction term ($E_{corr}$) using the SCPC scheme.}
    \label{SI_Eformcaxis}
\end{figure*}

\begin{figure*}[ht]
    \centering
    \includegraphics[width=0.4\textwidth]{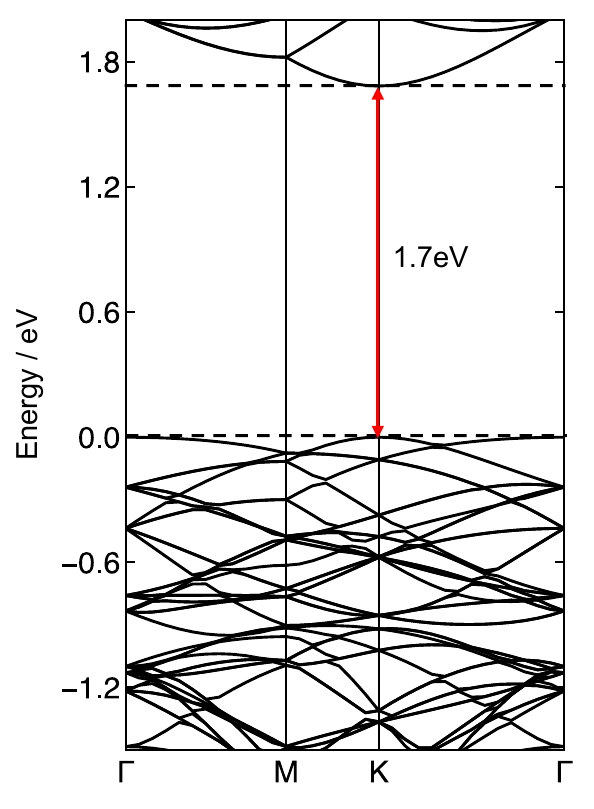}
    \caption{Band structure of $5 \times 5$ supercell of pristine \defect{MoS}{2} monolayer. The Fermi energy is set at $E_F = 0\ eV$. The dotted lines highlight VBM and CBM. The calculated band gap $E_g$ is $1.7\ eV$.}
    \label{SI_Pristine BS}
\end{figure*}

\begin{figure*}[ht]
    \centering
    \includegraphics[width=0.9\textwidth]{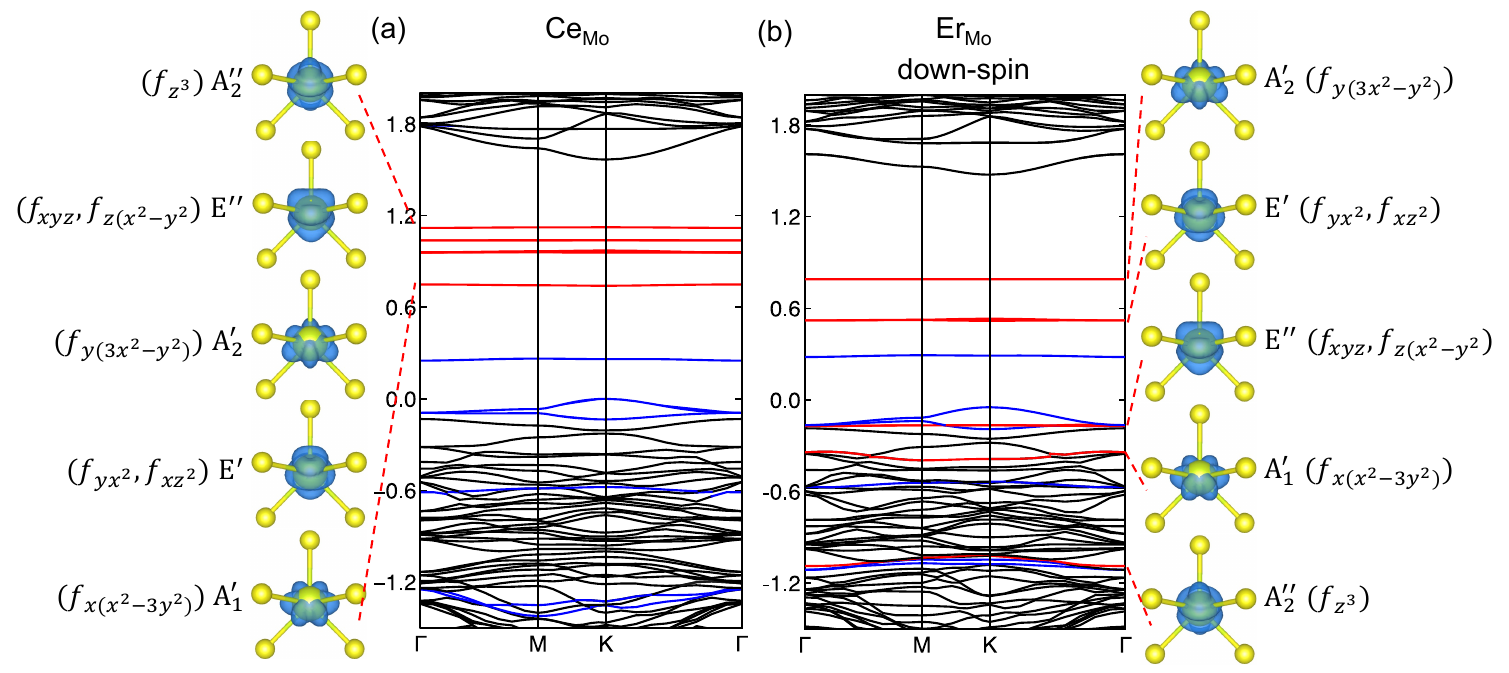}
    \caption{Band structures and charge distributions of $f$-orbitals of (a) \defect{Ce}{Mo} and (b) \defect{Er}{Mo} (down-spin). The red and blue bands show $f$-orbitals and defect states not originating from $f$-orbitals, respectively. The Fermi energy is set at $E_F = 0\ eV$. The charge density isosurface value is set to 0.007. Seven $f$-orbitals are divided into five eigenstates by $D_{3h}$ point group symmetry.}
    \label{SI_forbital}
\end{figure*}

\begin{figure*}[ht]
    \centering
    \includegraphics[width=0.9\textwidth]{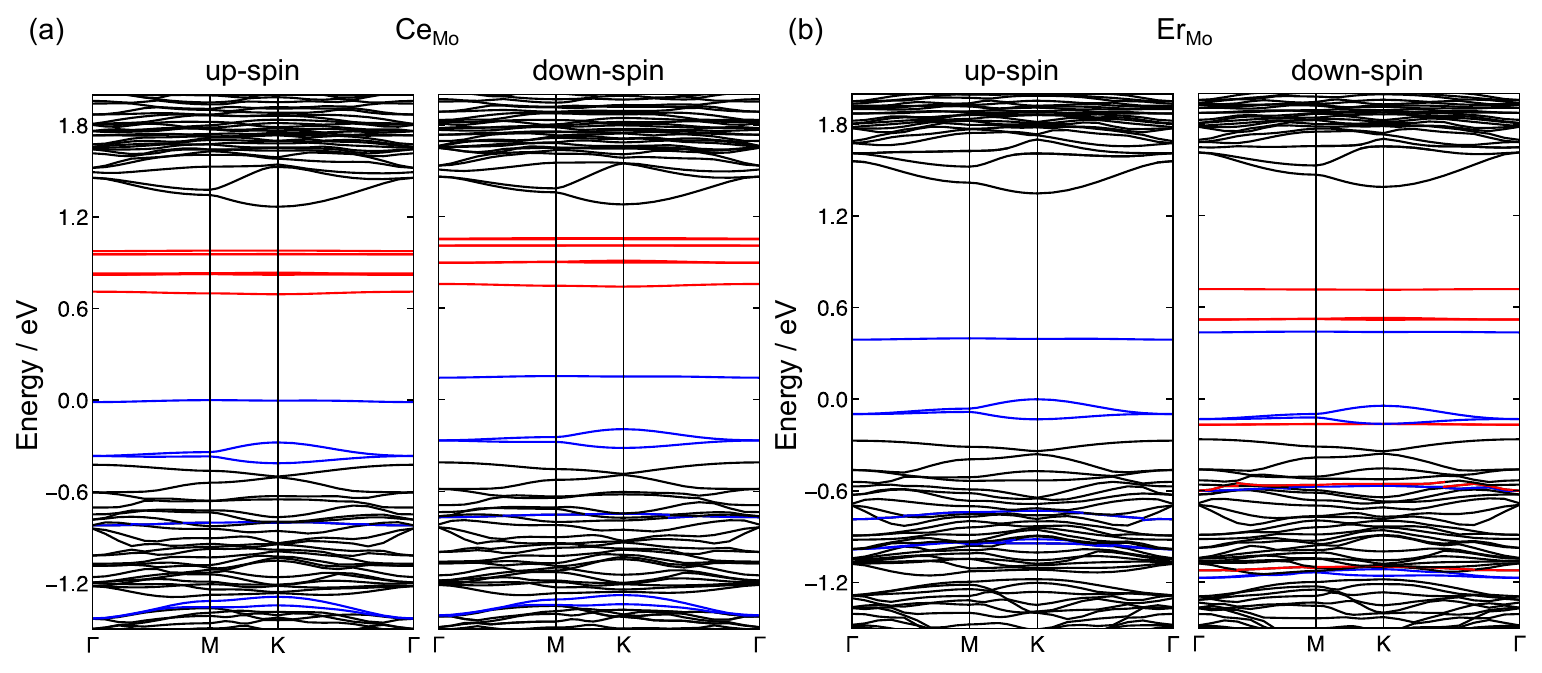}
    \caption{Band structures of negatively charged (a) \defect{Ce}{Mo} and (b) \defect{Er}{Mo}. The red and blue bands show $f$-orbitals and defect states not originating from $f$-orbitals, respectively. The Fermi energy is set at $E_F = 0\ eV$. Note that an additional electron does not contribute to $f$-orbitals, but defect states.}
    \label{SI_LnMo-}
\end{figure*}

\begin{figure*}[ht]
    \centering
    \includegraphics[width=0.9\textwidth]{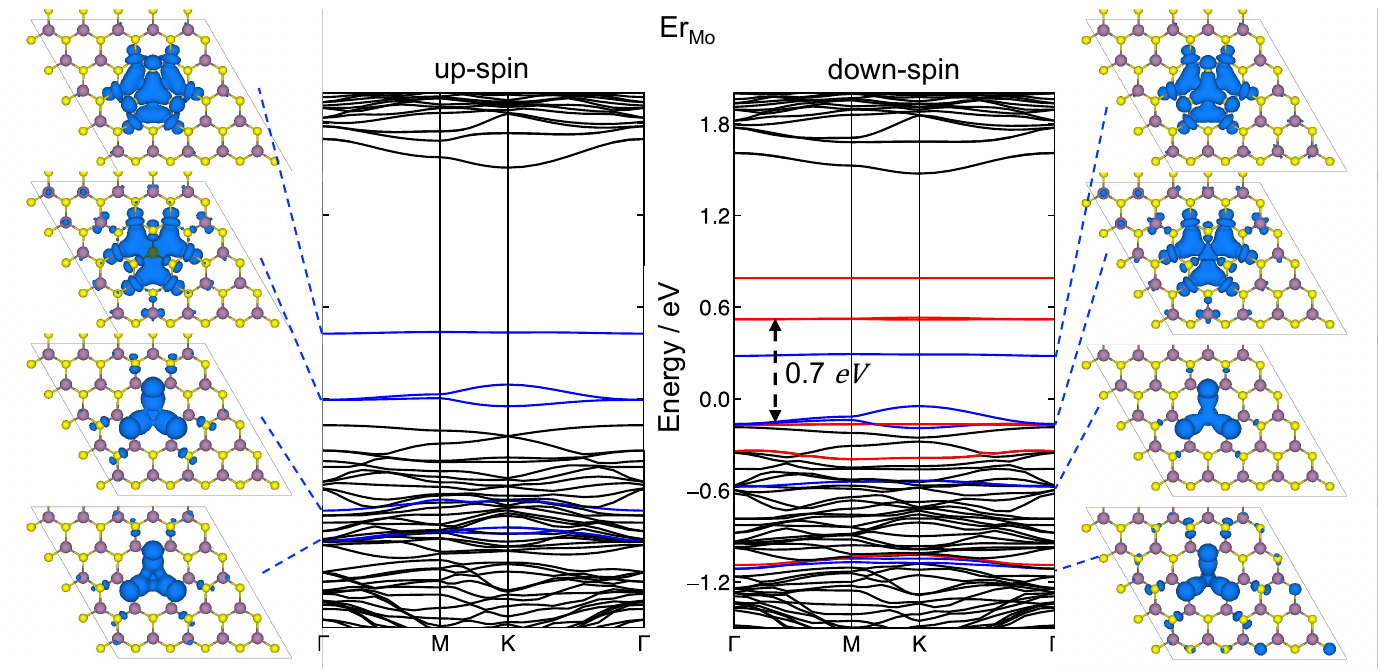}
    \caption{Band structures and charge distributions of defect states (blue bands) of \defect{Er}{Mo}. The red and blue bands show $f$-orbitals and defect states not originating from $f$-orbitals, respectively The Fermi energy is set at $E_F = 0\ eV$. The charge isosurface value is 0.001. The features and ordering of defect states are similar to those of \defect{Ce}{Mo}.}
    \label{SI_ErMoPCHGD}
\end{figure*}

\begin{figure*}[ht]
    \centering
    \includegraphics[width=\textwidth]{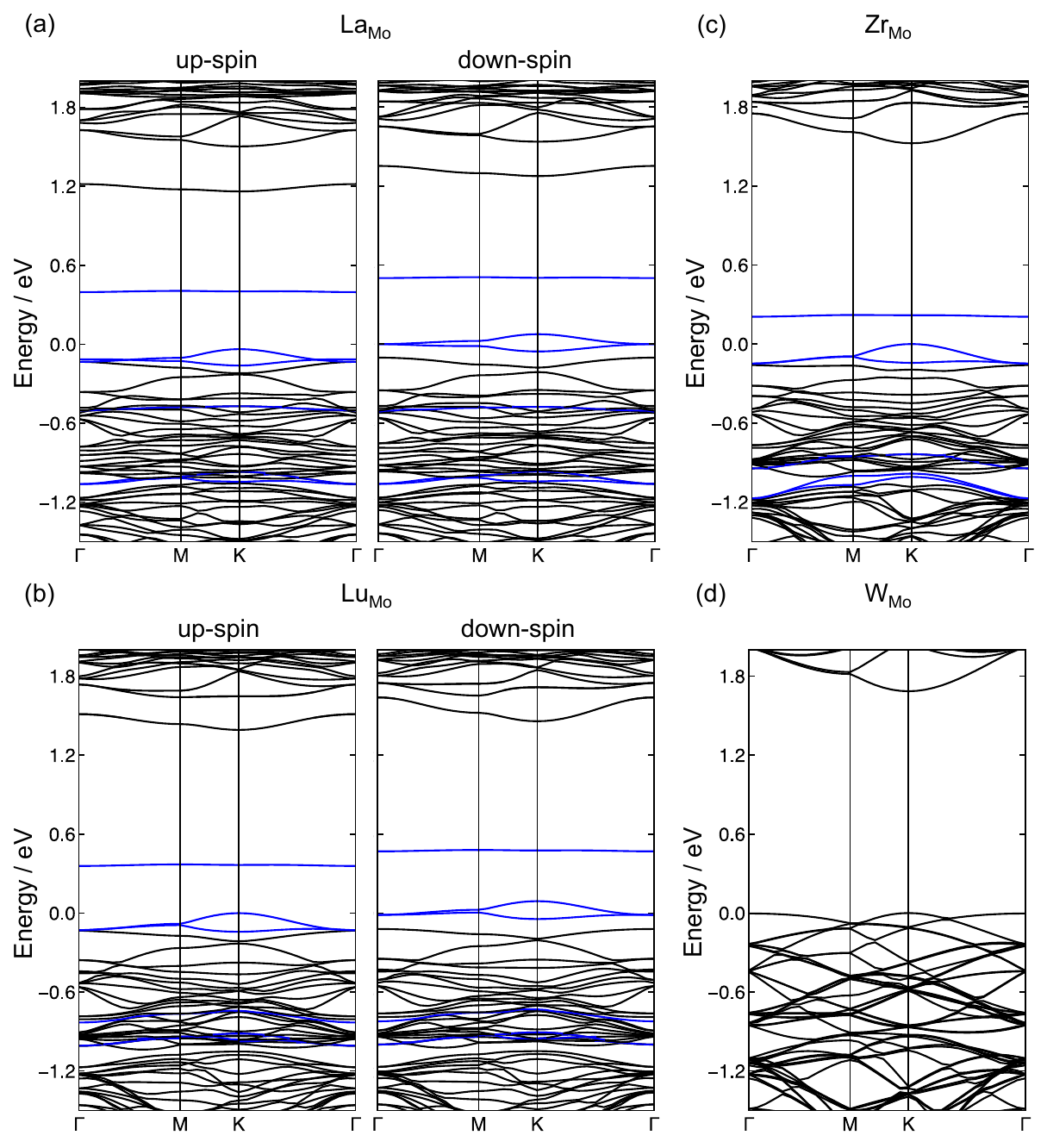}
    \caption{Band structures of (a) \defect{La}{Mo}, (b) \defect{Lu}{Mo}, (c) \defect{Zr}{Mo}, and (d) \defect{W}{Mo}. The blue bands show defect states not originating from $f$-orbitals. The Fermi energy is set at $E_F = 0\ eV$. Note that La$^{3+}$ ([Xe]), Lu$^{3+}$ ([Xe]$4f^{14}$), and Zr$^{4+}$ ([Kr]) possess a $d^0$ configuration while W$^{4+}$ ([Xe]$d^2$) possesses a $d^02$ configuration. This shows that an orbital occupation mismatch between Mo and the substituent atom leads to the emergence of defect states.}
    \label{SI_LaMoWMo}
\end{figure*}

\end{document}